\documentclass[]{aa}
\usepackage{psfig}
\usepackage{txfonts}
\begin{document}

\title{Simulating star formation in molecular cloud cores I.  The
influence of low levels of turbulence on fragmentation and multiplicity}
\titlerunning{Simulating star formation in molecular cloud cores I}

\author{S.\,P.\,Goodwin, A.\,P.\,Whitworth \and D.\,Ward-Thompson}
\authorrunning{S.\,P.\,Goodwin, A.\,P.\,Whitworth \& D\,Ward-Thompson}

\offprints{Simon.Goodwin@astro.cf.ac.uk}

\institute{Dept. of Physics \& Astronomy, Cardiff University, 5 The 
Parade, Cardiff, CF24 3YB, UK}

\date{}

\abstract{
We present the results of an ensemble of simulations of the collapse 
and fragmentation of dense star-forming cores. We show that even with 
very low levels of turbulence the outcome is usually a binary, or 
higher-order multiple, system.

\hspace{0.3cm} We take as the initial conditions for these simulations a typical 
low-mass core, based on the average properties of a large sample 
of observed cores. All the simulated cores start with a mass of 
$M_{\rm total} = 5.4 M_{\odot}$, a flattened central density profile, 
a ratio of thermal to gravitational energy $\alpha_{\rm therm} = 0.45$ 
and a ratio of turbulent to gravitational energy $\alpha_{\rm turb} 
= 0.05\,$. Even this low level of turbulence -- much lower than 
in most previous simulations -- is sufficient to produce multiple 
star formation in 80\% of the cores; the mean number of stars and
brown dwarfs formed 
from a single core is 4.55, and the maximum is 10. At the outset, the 
cores have no large-scale rotation.  The only difference between each 
individual simulation is the detailed structure of the turbulent 
velocity field.

\hspace{0.3cm} The multiple systems formed in the simulations 
have properties consistent with 
observed multiple systems.  Dynamical evolution tends preferentially to 
eject lower mass stars and brown dwarves whilst hardening the 
remaining binaries so that the median semi-major axis of binaries formed 
is $\sim 30$ au.  Ejected objects are usually single low-mass
stars and brown dwarfs, yielding a strong correlation between 
mass and multiplicity.   Brown dwarves are ejected with a higher average 
velocity than stars, and over time this should lead to mass segregation 
in the parent cluster.  Our simulations suggest a natural mechanism 
for forming binary 
stars that does not require large-scale rotation, capture, 
or large amounts of turbulence.

}

\maketitle

\section{Introduction}

There is now strong, almost irrefutable evidence that most stars are 
formed in binary or higher multiple systems. First, most mature field 
stars are observed to be in 
binary, or higher multiple, systems (Duquennoy \& Mayor 1991; Tokovinin 
\& Smekhov 2002). Second, the multiplicity of young pre-Main Sequence 
stars is if anything even higher than for mature field stars (Reipurth \& 
Zinnecker 1993; Reipurth 2000; Patience et al. 2002). Third, the ages 
of the components of young multiple systems are very similar (White \& 
Ghez 2001). Fourth, the properties of pre-Main-Sequence and Main-Sequence 
binaries (mass-ratios, separations and eccentricities) are very similar 
(Mathieu 1994; Bodenheimer et al 2000; Mathieu et al 2000). Fifth, the 
alternative -- namely that most stars form singly, and then pair up later, 
by dynamical and/or tidal interactions --  is too inefficient to produce 
the high observed frequency of multiple systems (Kroupa \& Burkert 2001). 
Thus a mechanism is required that produces multiple fragmentation within 
the majority of star forming cores.

There are two possibilities here. The first possibility is that 
fragmentation of a core is caused by its large-scale ordered 
rotation (see e.g. Bodenheimer et al. 2000). Large-scale ordered 
rotation can be parametrised by the initial ratio of rotational to 
gravitational energy $\beta$, and many numerical and semi-analytic 
studies have sought to identify the conditions required for 
fragmentation of a prestellar core, in terms of its ratio of thermal
to gravitational energy $\alpha_{\rm therm}$, 
$\beta$, other initial conditions such as shape and density profile, and its 
constitutive physics, in particular the equation of state (see 
Hennebelle et al. 2003, for a recent review). Observational estimates 
of $\beta$ in 
prestellar cores range from $10^{-4}$ to 0.07 (e.g. Caselli et al. 2002), 
and the higher values appear to be sufficient to promote fragmentation. 
However, many prestellar cores show no discernible evidence for 
rotation (e.g. Jessop \& Ward-Thompson 2001), suggesting that 
large-scale ordered rotation is not the sole cause of fragmentation.

The second possibility is that fragmentation of a core is caused by 
turbulence, i.e. by more disordered small-scale bulk velocity fields, 
which may conspire to create local angular momentum, but do not constitute 
an overall rotational velocity field in the core as a whole. 
Observations of prestellar cores (e.g. Caselli et al. 2002) detect 
complex motions suggestive of turbulence.

Turbulence is a simple way in which to introduce velocity and density 
inhomogeneities that may then seed multiple fragmentation. For instance, 
in the context of molecular cloud evolution, the work of 
Elmegreen (1997), Padoan et al. (1997), Klessen \& 
Burkert (2001), and  Padoan \& Nordlund (2002) (see also the reviews
of Mac Low \& Klessen 2003; Larson 2003)  suggests that turbulence 
plays a key r\^ole in forming 
dense cores. The distribution and characteristics of the  
cores that are formed depends critically on the scale-length on which 
the turbulence is driven (see Klessen 2003, and references therein). 
Moreover, many of the cores which form are transient structures, and 
only a small subset of them become dense enough to condense into 
stars (Ballesteros-Paredes et al. 2003; 
Klessen et al. 2003). It is with this subset, the {\it dense cores}, 
that this paper is concerned.

In the context of core evolution, the work of Klein et al. 
(2001), Bate et al. (2002a,b; 2003), Klein et al. (2003), 
and Bonnell, Bate \& Vine (2003) indicates that turbulence can be very 
effective in promoting the fragmentation of collapsing cores; the 
semi-analytic work of Fisher (2003) suggests that turbulence is the 
underlying cause of the distribution of binary periods. However, 
most previous simulations of collapsing cores have started from high 
initial levels of turbulence. Observations indicate that isolated 
prestellar cores actually have rather narrow line widths (e.g. 
Beichman et al. 1986; Jijina et al. 1999), suggesting low levels of 
turbulence.

The problem addressed in this paper is therefore to reconcile the 
low levels of turbulence observed in many prestellar cores with the 
expectation that they will spawn multiple stellar systems. By means of
numerical simulations, can low 
levels of turbulence produce multiple fragmentation? We also examine 
the statistical properties of the resultant stars and multiple 
systems formed in the simulations. In Section 2 we 
describe observations of dense molecular cores and the initial 
conditions we infer from these observations.  In Section 3 we 
outline the code and numerical aspects of the simulations, and in 
Section 4 we present the results of our simulations. We discuss the
results in Section 5, and present the conclusions in Section 6.

Star formation in the presence of turbulence is expected to be a 
stochastic process (Larson 2001), and therefore a statistical ensemble of 
simulations must be performed in order to compare meaningfully with 
observations. In this paper we present an ensemble of 20 simulations 
of star formation in a realistic (i.e. observationally motivated), 
dense core.

\section{Empirical initial conditions}

Here we outline the observational constraints on the properties 
of molecular cores that are on the verge of protostellar collapse
and describe the initial conditions that we adopt.

\subsection{Observational constraints}

Molecular cores can be divided into those that appear to have
formed protostars in their centres -- i.e. those associated with 
IRAS sources -- and those that do not (e.g. Beichman et al. 1986). 
The latter show no evidence for outflows, and are usually referred to as 
starless cores. Starless cores typically have masses of a few 
$M_{\odot}$, volume densities of $10^3$ to $10^4\,{\rm cm}^{-3}$, 
radii $\sim\,0.1\,{\rm pc}$ and are approximately isothermal, with 
temperatures $\sim\,10\,{\rm K}$ (e.g. Jijina et al. 1999).

Those starless cores that are most centrally condensed, and hence
presumably closest to protostellar collapse, are known as pre-stellar
cores (originally pre-protostellar cores -- Ward-Thompson et al. 1994).
Their density profiles are approximately flat within a 
central region of a few thousand au, and then fall as $r^{-2}$ before 
steepening even further to $r^{-4}$ or $r^{-5}$ at their edges
(e.g. Ward-Thompson et al. 1994; Andr\'e et al. 1996; 
Ward-Thompson et al. 1999). They finally become 
indistinguishable from the background molecular cloud at their edges
(see Andr\'{e} et al. (2000) for a review of 
core properties).

Observations of cores show that they often have a significant 
non-thermal contribution to their line widths, which can be 
attributed to turbulence. Larson (1981) has shown that on average 
the line-of-sight velocity dispersions, $\sigma$, of molecular 
clouds and cores are related to their tangential linear sizes, $L$, 
by a relation of the form
\begin{equation} \label{LARSON1}
\sigma = 1.10\,{\rm km}\,{\rm s}^{-1}\,
\left(L/{\rm pc}\right)^{0.38}\,,
\end{equation}
although later studies have suggested that the exponent should be 
higher than 0.38 (e.g. Myers 1983) and depends on the presence of 
embedded IRAS sources (Jijina et al. 1999). Equation (\ref{LARSON1}) 
is normally interpreted as evidence for turbulence within molecular 
clouds (see Larson 2003 for a review).  

A key parameter characterising the level of turbulence in a core is 
the ratio of turbulent to gravitational energy, $\alpha_{\rm turb}$. 
Figure~\ref{fig:nonthermal} shows a plot of $\alpha_{\rm turb}$ against 
core mass for the starless cores in the Jijina et al. (1999) catalogue. 
We note that there is some difficulty in accurately determining the masses 
of cores, and hence their gravitational energies, but we believe 
that the points plotted in Fig.~\ref{fig:nonthermal} are accurate to 
within a factor of two or better, and representative of dense cores 
which are destined to form stars.

Figure~\ref{fig:nonthermal} shows that the average $\alpha_{\rm turb}$ is low; 
almost all of the observed dense cores have $\alpha_{\rm turb}<0.5$, and most have $0<\alpha_{\rm turb}<0.3$. The star symbol in the lower half of 
Fig.~\ref{fig:nonthermal} represents the parameters (mass and turbulence) 
of the cores we have simulated here, chosen to be representative of dense 
cores with low levels of turbulence. The open circle in the upper right 
of Fig.~\ref{fig:nonthermal} represents the parameters of the core 
simulated by Bate et al. (2002a,b, 2003).

\subsection{Initial conditions}

\begin{figure*}
\centerline{\psfig{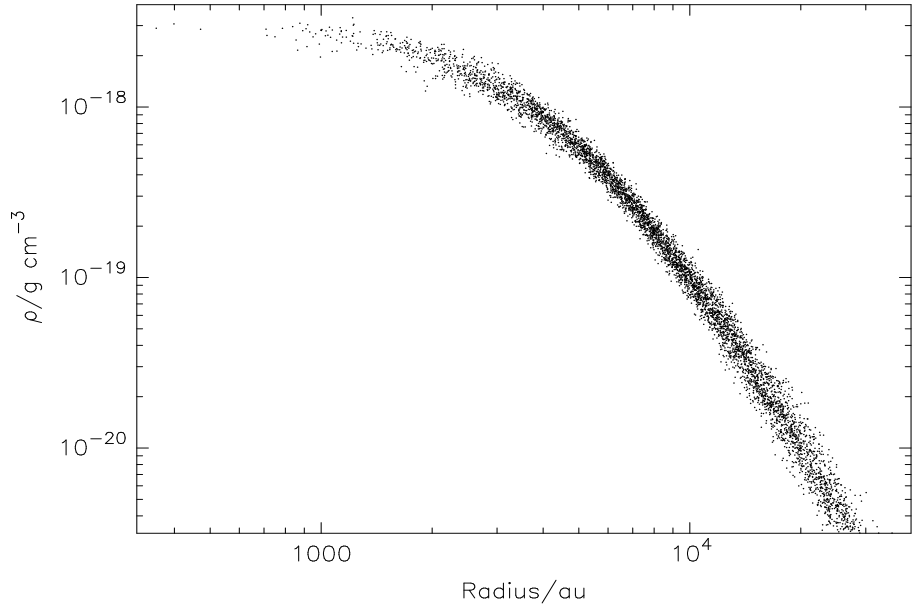}}
\caption{The density profile of a prestellar core with a flat inner
region (kernel) and a $1/r^4$ decline in the outer envelope.}
\label{fig:profile}
\end{figure*}

\begin{figure*}
\centerline{\psfig{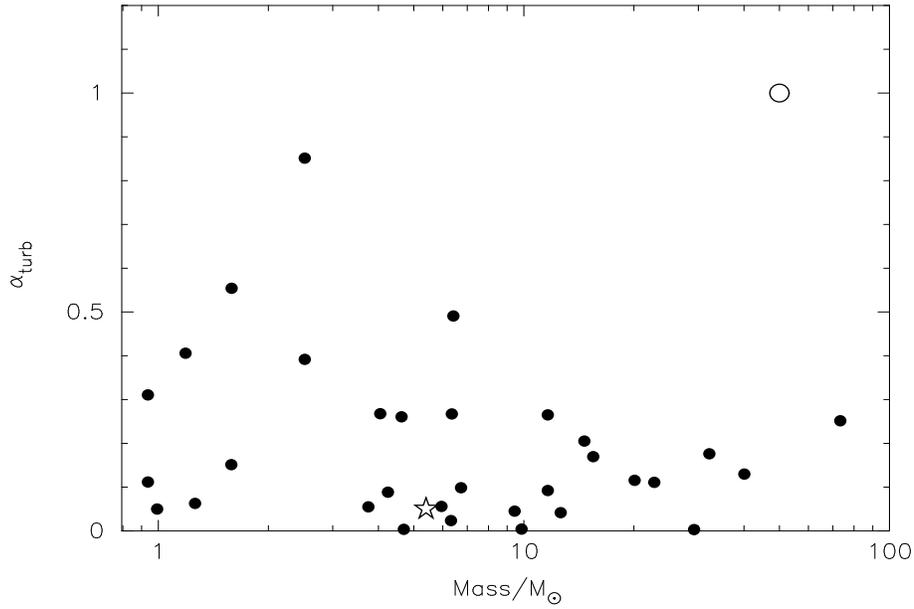}}
\caption{The ratio of turbulent to gravitational energy 
$\alpha_{\rm turb}$ plotted against mass, for observed starless cores 
from Jijina et al. (1999). Also marked are the point in parameter 
space treated by our simulations (the star symbol) and that treated 
by Bate et al. (2002a,b, 2003; the open circle).}
\label{fig:nonthermal}
\end{figure*}

A Plummer-like profile of the form 
\begin{equation} \label{PLUMMER}
\rho(r) = \frac{\rho_{\rm kernel}}{(1 + (r/R_{\rm kernel})^2)^2}
\end{equation}
gives a good fit to the observed density profiles of dense cores 
(see Fig.~\ref{fig:profile}). $\rho_{\rm kernel}$ is the central density 
(here $3 \times 10^{-18}\,{\rm g}\,{\rm cm}^{-3}$). $R_{\rm kernel}$
is the radius 
of the central region of the core -- hereafter the kernel -- in which 
the density is approximately uniform (here $5,000\,{\rm au}$). In 
the outer envelope of the core the density falls off as $r^{-4}$, 
and the outer boundary of the core is at $50,000\,{\rm au}\,$. The 
total mass of the core is $5.4 M_{\odot}$, of which $\sim 2 M_\odot$ 
is in the kernel. The core is initially isothermal, with temperature 
$T = 10\,{\rm K}\,$, and hence $\alpha_{\rm therm} = 0.45\,$.

\begin{table}
\caption{Initial conditions for low-turbulence cores.}
\label{tab:binary}
\begin{tabular}{lc}
\hline
& \\
Parameter & Value \\ \hline
& \\
Central density & 3 $\times$ 10$^{-18}$ g cm$^{-3}$ \\
Kernel radius & $5,000\,{\rm au}$ \\
Maximum radius & $50,000\,{\rm au}$ \\
Temperature & $10\,{\rm K}$ \\
Mass & $5.41 M_{\odot}$ \\
Thermal virial ratio & 0.45 \\
Turbulent virial ratio & 0 or 0.05 \\
& \\
\hline
\end{tabular}
\end{table}

A divergence-free Gaussian random velocity field is superimposed on 
the core to simulate turbulence (cf. Bate et al. 2002a,b, 2003; 
Bonnell et al. 2003).  The power spectrum of the turbulence 
is set to $P(k) \propto k^{-4}$, which mimics the Larson scaling 
relation of Eqn. (\ref{LARSON1}) (Burkert \& Bodenheimer 2000). 
The level of turbulent energy is set to $\alpha_{\rm turb} = 0.05\,$. 
Hence the core is initially globally virialised, in the sense that
$\alpha_{\rm therm} + \alpha_{\rm turb} = 0.5$, although it is not 
in detailed hydrostatic balance. In a subsequent paper we will 
examine the effect of varying the level of turbulent energy.

We present an ensemble of 20 simulations with these initial conditions. 
The only difference between individual simulations is that the random 
number seed for the turbulent velocity field is changed - thereby 
changing the detailed structure of the velocity field, but not its 
overall magnitude. In addition we have run 10 simulations with no 
turbulence for the purpose of comparison; in each of these simulations 
the positional random number seed is changed so that the resulting 
Poisson noise in the initial particle positions is different.

The simulations are allowed to run for 0.3 Myr. 
After 0.2 Myr most of the dynamical evolution is finished, although 
accretion is still on-going. At around 0.2 to $0.3\,{\rm Myr}$, feedback 
from the newly-formed stars (through jets and outflows) is likely to 
have a significant effect on the evolution of the remaining gas, by 
dispersing the core envelope and terminating further accretion. 
However, this is not included in the simulations reported here.

\section{Computational details}

We use the Smoothed Particle Hydrodynamics (SPH) code  
{\sc dragon}, which is based on the original Cardiff group code, as 
described in detail by Turner et al. (1995), but has recently 
been rewritten from scratch and greatly optimised. SPH is a 
particle-based, Lagrangian scheme in which the particles represent 
sampling points in the gas. It is ideal for simulations of star 
formation, because as the density increases the resolution also 
increases due to the concentration of particles in dense regions. 
Our SPH implementation is standard (see Monaghan 1992). It uses a 
variable smoothing length with the constraint that the number of 
neighbours is $N_{\rm neib} \sim 50$.  An octal tree is used to 
calculate the neighbour lists and gravitational accelerations, 
which are kernel-softened using the particle smoothing 
lengths $h$.  Artificial viscosity is included in converging regions 
with $\alpha_{\rm v} = 1$ and $\beta_{\rm v} = 2$. Multiple 
particle time-steps are invoked.

\subsection{Equation of state and the opacity limit for fragmentation}

During the early stages of collapse, the rate of compressional heating
is low and the gas is able to cool radiatively, either by molecular 
line emission, or, when $\rho > 10^{-19}$ g cm$^{-3}$, by coupling 
thermally to the dust.  As a consequence, the gas is approximately 
isothermal, with $P \propto \rho$.  However, eventually the rate of 
compressional heating becomes so high (due to the acceleration 
of the collapse) and the rate of radiative cooling becomes so low 
(due to the increasing column density and hence increasing dust optical 
depth), that the gas switches to being approximately adiabatic, with 
$P \propto \rho^{5/3}$. For cores with mass in the range 1 to 
$10 M_{\odot}$ and initial temperature $T \sim 10\,{\rm K}$, the switch from isothermality to adiabaticity occurs at a critical density 
$\rho_{\rm crit} \sim 10^{-13}\,{\rm g}\,{\rm cm}^{-3}$ (Larson 1969; 
Tohline 1982; Masunaga \& Inutsuka 2000).  We reproduce this behaviour 
using a barotropic equation of state:
\begin{equation} \label{EofS}
\frac{P(\rho)}{\rho} \equiv c_s^2(\rho) = c_0^2 \left[ 1 + \left
( \frac{\rho}{\rho_{\rm crit}} \right)^{2/3} \right]
\label{eqn:eos}
\end{equation}
Here $P$ is the pressure, $\rho$ is the density, $c_s$ is the general 
isothermal sound speed, and $c_0 = 0.19\,{\rm km}\,{\rm s}^{-1}$ is 
the isothermal sound speed in the low-density gas, which is presumed 
to be molecular hydrogen at $10\,{\rm K}\,$.

\subsection{Resolution and the Jeans mass}

The Jeans mass is given by
\begin{equation} \label{eqn:jeansmass}
M_J(\rho) \simeq \frac{6 c_s^3(\rho)}{G^{3/2} \rho^{1/2}}
\end{equation}
where $G$ is the gravitational constant. In the low-density isothermal 
phase, the Jeans mass decreases as the gas density increases, and in 
the high-density adiabatic phase, the Jeans mass increases with 
increasing density. There is therefore a minimum Jeans mass,
\begin{equation}
M_{\rm crit} \label{Mcrit}
\simeq \frac{17 c_0^3}{G^{3/2} \rho_{\rm crit}^{1/2}} 
\simeq 0.01 M_\odot 
\end{equation}
at $\rho_{\rm crit}$.

A number of studies has shown that it is essential to resolve the 
Jeans mass in simulations of fragmentation (e.g. Bate \& Burkert 1997; 
Whitworth 1998; Truelove et al 1998; Sigalotti \& Klapp 2001; Kitsionas 
\& Whitworth 2002). This requirement is usually referred to as the Jeans 
Condition. Violation of the Jeans Condition can either result in 
artificial fragmentation, or the suppression of real fragmentation 
(Bonnell 2003). Since the minimum mass that can be resolved by 
{\sc dragon} is $\sim {\cal N}_{\rm neib} m$, where $m$ is the mass 
of a single SPH particle, the Jeans Condition becomes an upper limit 
on $m$:
\begin{equation}
m \lesssim m_{\rm max} 
\simeq \frac{M_{\rm crit}}{{\cal N}_{\rm neib}} 
\simeq 0.0002 M_\odot \,.
\end{equation}
Therefore if we are to model a $5.4 M_\odot$ core, we need at least 
25,000 equal-mass SPH particles, and this is the number we have used for 
the majority of the simulations reported here, in order to be able to 
perform a statistically significant ensemble of simulations.

In order to test for convergence, we have repeated a simulation which 
(when run with 25,000 particles) only produces a single protostar, 
using 50,000 and 100,000 particles. All three simulations have the 
same turbulent velocity field, but different initial particle noise, 
and all three simulations produce only one object. At no point do the 
higher resolution simulations show any tendency to produce a second 
protostar. After formation of the primary (and only) protostar, the
maximum density peaks at $\sim 10^{-12}\,{\rm g}\,{\rm cm}^{-3}$, and 
this peak is always in the immediate vicinity of the primary protostar. 
Other simulations have been tested for convergence in the same way. 
In no case do we find a significant difference in the evolution.

\subsection{Sink particles}

As the density of a collapsing core increases, simulations become 
very computationally expensive, because the timestep required 
accurately to integrate particle trajectories in the densest regions 
becomes extremely short. In order to alleviate this problem, regions 
whose density greatly exceeds the critical density (i.e. $\rho 
> \rho_{\rm sink} = 100\,\rho_{\rm crit}$) are replaced by sink 
particles of radius $5\,{\rm au}$ (see Bate et al. 1995 
for a full description of sink particles).

Sink particles are effectively black boxes within which a fragment 
evolves and develops into a protostar. SPH particles which enter the sink 
radius, and are bound to the sink, are merged with the sink, and its 
mass, momentum and centre of mass are adjusted accordingly. Sinks 
interact solely through gravity, which is kernel-softened using the 
sink radius. The maximum timestep for a sink is set to be 0.05 yrs. 
This allows orbits to be integrated accurately down to separations 
of $\sim 10\,{\rm au}\,$ which is well below the peak in the binary 
separation distribution at $\sim 30\,{\rm au}\,$.

Close encounters between protostars are a very important part of the 
evolution of newly-formed multiple systems. In the simulations presented 
here, sinks are not allowed to merge. In the period when sinks represent 
extended pressure-supported fragments this is not a reliable 
representation of their behaviour, as encounters are likely to result in 
mergers. However, the lifetime of the extended pressure-supported 
phase is only $\sim 2 \times 10^4\,{\rm yrs}$ (Masunaga \& Inutsuka 2000). 
After this time, the dissociation of molecular hydrogen allows protostars 
to collapse to stellar densities, and collisions would then be very rare 
in the environments we simulate -- although in denser environments collisions 
may provide a mechanism for forming very massive stars (Bonnell \& Bate 
2002).

\begin{figure*}
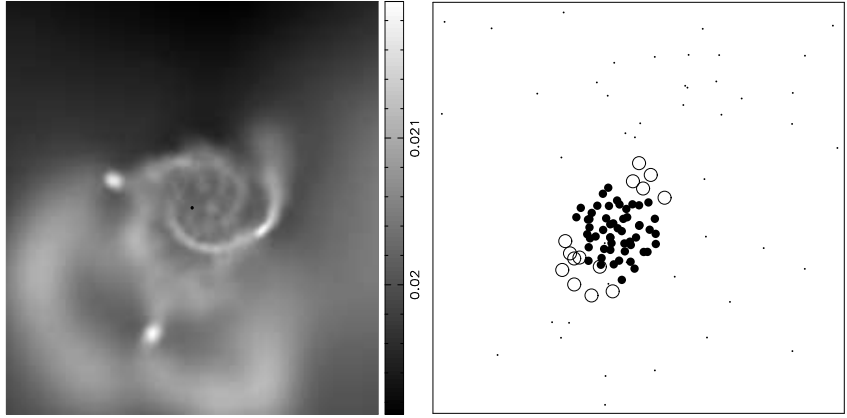

\centerline{\psfig{figure=GWWT_fig3a.ps,height=5.5cm,width=5.5cm,angle=270}\,\,\,\psfig{figure=GWWT_fig3b.ps,height=5.5cm,width=5.5cm,angle=270}}
\caption{(a) A column-density plot of the region around a sink. The
view is $500\,{\rm au}$ across, and the grey-scale bar indicates the 
column density in ${\rm g}\,{\rm cm}^{-3}$. The black dot at the
centre indicates the position of the sink. (b) A close-up, 
$100\,{\rm au}$ across, 
of the region in the bottom middle of (a) that is about to form another 
sink. This shows that the particles about to form the sink (large 
points) have collapsed into a spherical region and are not strung out 
along a filament. The elongation of the region is caused by particles 
which are accreted shortly after sink formation (open circles).}
\label{fig:nostring}
\end{figure*}

A significant concern in low-resolution simulations is that a filament 
which should fragment into more than one condensation will just form a single 
sink. However, Fig.~\ref{fig:nostring} demonstrates that, by setting 
$\rho_{\rm sink} = 100 \rho_{\rm crit}$, we ensure that an unstable 
region that will form a sink has condensed to an approximately 
spherical object 
prior to sink creation (see also Fig.~\ref{fig:further}). This
shows that the creation of sink particles is not suppressing the 
fragmentation of filaments.

\section{Results}

The most significant results of this investigation can be summarised 
succinctly. First, a small level of turbulence induces the 
formation of multiple systems in 80\% of cores. Second, the number 
and properties of the objects formed in an individual core depends, 
in a chaotic way, on the details of the initial turbulent velocity 
field, ranging from one object to ten with an average of 4.55 objects 
per core. Third, the properties of the binary and multiple systems 
that form are compatible with observation.

Henceforth we use the term `objects' to refer to sinks formed in the
simulations, and then more specifically `stars' for objects with
masses greater than $0.08 M_{\odot}$ and `brown dwarfs' for objects of
lower mass.

\subsection{No turbulence}

To provide a reference for the turbulent simulations, we conduct 10
runs with no turbulence. Here the only difference between individual 
simulations is the random number used to seed the initial positions 
of the SPH particles. In all cores with no turbulence, only one object 
forms, always a star, and always within $100\,{\rm au}$ of the centre 
of mass. This is exactly what would be expected from the collapse of an 
initially static and slightly sub-virial core.  

\begin{figure*}
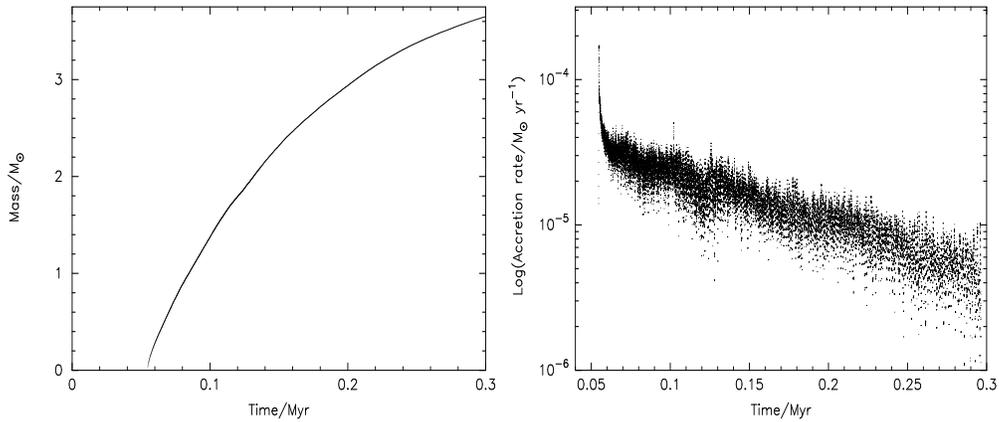

\centerline{\psfig{figure=GWWT_fig4a.ps,height=5.5cm,width=6.5cm,angle=270}\,\,\,\psfig{figure=GWWT_fig4b.ps,height=5.5cm,width=6.5cm,angle=270}}
\caption{Left: the mass of the single star that forms in a core 
with no turbulence, as a function of time. Right: the accretion 
rate as a function of time for the same star. The accretion rate 
is in good agreement with the semi-analytic model of Whitworth \& 
Ward-Thompson (2001).}
\label{fig:zero_accrete}
\end{figure*}

In all cases, the accretion histories of the stars are indistinguishable, 
An example is given in Fig.~\ref{fig:zero_accrete}, which shows the mass 
and accretion rate of the star as a function of time. The accretion rate 
is initially high ($\sim 10^{-4} M_{\odot}\,{\rm yr}^{-1}$) and then 
drops rapidly over the course of the next $\sim 10^5\,{\rm yrs}\,$. The 
high initial accretion rate is due to material from the uniform density 
kernel of the core, which, having approximately uniform density, 
undergoes nearly freefall collapse. The accretion rate subsequently 
decreases as material from the lower density envelope falls inwards.

The evolution of these zero-turbulence cores follows closely the semi-analytic 
model of Whitworth \& Ward-Thompson (2001) who assumed negligible internal 
pressure and hence free-fall collapse. The central density rises to the 
sink formation threshold of $10^{-11}\,{\rm g}\,{\rm cm}^{-3}$ after 
around $0.05\,{\rm Myr}$. Subsequent accretion 
(Fig.~\ref{fig:zero_accrete} (b)) then follows closely the pattern in Fig. 3 
of Whitworth \& Ward-Thompson (2000). The accretion rate at very early
times is slightly in excess of the Whitworth \& Ward-Thompson (2001)
value due to the SPH particles having finite mass.

\subsection{Turbulent cores}

In order to evaluate the effect of turbulence we have conducted an 
ensemble of 20 simulations with $\alpha_{\rm turb} = 0.05$. In 
Table~\ref{tab:summary} we summarise the results of these simulations, 
$0.3\,{\rm Myr}$ after their start.  In 80\% of simulations, turbulence 
results in the formation of multiple objects, and the number of objects 
that forms depends only on the detailed structure of the initial 
turbulent velocity field (i.e. the random-number seed that generates 
the turbulence). A total of 91 objects is formed in the 20 simulations, 
75 stars and 16 brown dwarfs, averaging 4.55 objects per core.  
Each core produces between 1 and 4 high mass ($>0.5 M_{\odot}$) 
stars, and between 0 and 8 low mass ($<0.5 M_{\odot}$) objects.  
All but one of the brown dwarfs form in clouds that produce 
$\geq 6$  objects.  Star formation mostly occurs between 0.05 
and $0.10\,{\rm Myr}\,$, and only Run A013 undergoes star formation 
after $0.2\,{\rm Myr}\,$, forming 3 very low mass stars and one 
brown dwarf between 0.26 and 0.28 Myr.

In the main text we will limit our discussion to the general  
statistical properties of the ensemble of simulations. In Appendix 
1 we briefly outline the evolutionary histories of the individual 
cores.

\begin{center}
\begin{table*}
\caption[]{A summary of the results from all 20 low-turbulence simulations. 
The first column gives the run identifier; the second column gives the 
total number of objects that form, $N_{\rm obj}$; the third column gives 
the total mass accreted by all of the objects in a core by 
$0.3\,{\rm Myr}\,$, $M^*_{\rm tot}$; and the fourth column gives the 
resulting star-formation efficiency, $\eta$. The fifth column summarises 
the multiplicity of the final system of objects; Run A013 has recently 
undergone a burst of star formation and the multiple system is very 
unstable at the end of the simulation. The sixth column gives the 
masses of individual objects, marked with a $^b$, $^t$ or $^q$ if that 
object is part of a binary, triple or quadruple, respectively.}
\label{tab:summary}
\begin{tabular}{lcccll}
&&&&&\\
\hline
&&&&&\\
Run & $N_{\rm obj}$ & $M^*_{\rm tot}/M_{\odot}$ & $\eta$ &
Multiplicity & Masses/$M_{\odot}$  \\
&&&&&\\
\hline
&&&&&\\
A011 & 7  & 2.94 & 0.54 & Triple system & 1.31$^t$, 0.61$^t$, 0.52$^t$, 0.27, 0.12, 0.063, 0.048   \\
A012 & 4  & 3.72 & 0.69 & Wide binary  & 2.32$^b$, 0.74$^b$, 0.48, 0.18  \\
A013 & 10 & 3.10 & 0.57 & Binary + triple$^{*}$ & 1.07$^b$, 0.66$^b$, 
0.43, 0.34, 0.17, 0.13$^t$, 0.10$^t$, 0.09$^t$, 0.076, 0.040 \\
A014 & 3  & 4.02 & 0.74 & Triple system & 1.63$^t$, 1.56$^t$, 0.83$^t$ \\
A015 & 2  & 3.69 & 0.68 & Wide binary  & 2.63$^b$, 1.06$^b$   \\
A016 & 3  & 3.61 & 0.67 & Wide binary  & 2.18$^b$, 1.40$^b$, 0.028  \\A017 & 6  & 3.75 & 0.69 & Triple system & 1.60$^t$, 1.16$^t$, 0.64$^t$, 
0.18, 0.12, 0.050  \\
A018 & 7  & 3.65 & 0.67 & Triple system & 1.09$^t$, 1.03$^t$, 0.69$^t$, 
0.58, 0.18, 0.045, 0.041 \\
A019 & 8  & 3.81 & 0.70 & Triple system & 1.27$^t$, 1.16$^t$, 0.69$^t$, 
0.39, 0.21, 0.044, 0.030, 0.025 \\
A020 & 1  & 3.63 & 0.67 & Single star & 3.63  \\
A021 & 1  & 3.69 & 0.68 & Single star & 3.69  \\
A022 & 4  & 4.01 & 0.74 & Double binary & 1.52$^q$, 0.91$^q$, 0.89$^q$, 
0.69$^q$ \\
A023 & 4  & 3.56 & 0.66 & Triple system & 1.43$^t$, 0.83$^t$, 0.70$^t$, 
0.60  \\
A024 & 5  & 3.55 & 0.66 & Close binary & 1.46$^b$, 1.28$^b$, 0.43, 
0.19, 0.18  \\
A025 & 8  & 3.47 & 0.64 & Wide binary & 1.43$^b$, 0.76$^b$, 0.51, 0.47, 
0.14, 0.064, 0.045, 0.039 \\
A026 & 7  & 3.94 & 0.73 & Triple system & 1.23$^t$, 1.03$^t$, 0.73, 
0.71$^t$, 0.11, 0.098, 0.027 \\
A027 & 2  & 3.67 & 0.68 & Wide binary & 3.19$^b$, 0.48$^b$ \\
A028 & 1  & 3.35 & 0.62 & Single star & 3.35 \\
A029 & 7  & 3.61 & 0.67 & Double binary & 1.20$^q$, 0.89$^q$, 
0.57, 0.51, 0.29$^q$, 0.11, 0.041$^q$ \\
A030 & 1  & 2.62 & 0.48 & Single star & 2.62 \\
&&&&&\\
\hline
\end{tabular}
\end{table*}
\end{center}

\subsection{The origin of multiple fragmentation}

The initial collapse of a core causes a significant over-density 
to form near the centre of the core. Usually this over-density is 
flattened, i.e. either layer-shaped due to the convergence of two 
large turbulent elements, or disc-shaped due to local spin angular 
momentum. This flattened region then becomes unstable and forms 
the first sink particle, hereafter the primary star. The initial 
location of the primary star is typically within $2,000\,{\rm au}$ 
of the centre of mass of the core; its exact position depends upon 
the details of the turbulent velocity field. 

The primary star always forms roughly one free-fall time 
($\sim 0.05\,{\rm Myrs}$) after the beginning of the simulation. The 
low level of turbulence invoked here is not sufficient to delay 
the collapse of the core significantly. This is in contrast to the 
high-turbulence simulation of Bate et al. (2002a,b, 2003), 
where turbulent energy causes the core to expand initially, 
and results in the formation of strong filaments.

The primary star grows rapidly in mass by accreting from the 
surrounding over-density. The inflow is anisotropic, being 
strongest in the plane of the flattened over-density, lumpy, 
and variable. Usually it also has net angular momentum 
relative to the primary star, and in this case a circum-primary 
disc forms. Due to the lumpy, variable accretion, these 
discs become unstable to spiral modes and fragment to produce 
secondary objects. The genesis of secondary objects is usually 
concentrated in a burst between 0.07 and $0.08\,{\rm Myr}$ after 
the start of the simulation (or 0.01 to $0.03\,{\rm Myr}$ after the primary 
star forms).

\begin{figure*}
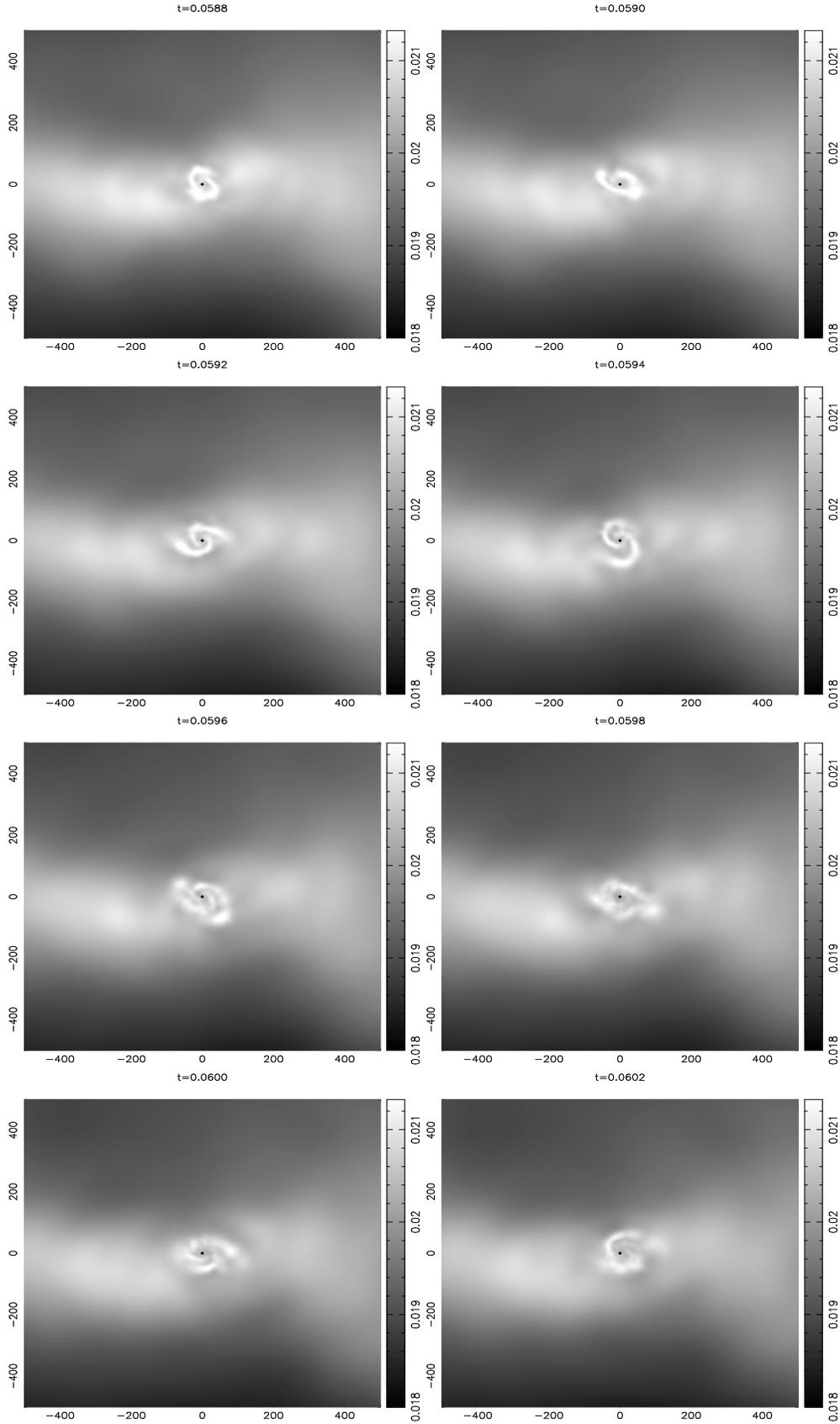

\centerline{\psfig{figure=GWWT_fig5a.ps,height=5.5cm,width=6.5cm,angle=270}\,\,\psfig{figure=GWWT_fig5b.ps,height=5.5cm,width=6.5cm,angle=270}}
\vspace{0.1cm}
\centerline{\psfig{figure=GWWT_fig5c.ps,height=5.5cm,width=6.5cm,angle=270}\,\,\psfig{figure=GWWT_fig5d.ps,height=5.5cm,width=6.5cm,angle=270}}
\vspace{0.1cm}
\centerline{\psfig{figure=GWWT_fig5e.ps,height=5.5cm,width=6.5cm,angle=270}\,\,\psfig{figure=GWWT_fig5f.ps,height=5.5cm,width=6.5cm,angle=270}}
\vspace{0.1cm}
\centerline{\psfig{figure=GWWT_fig5g.ps,height=5.5cm,width=6.5cm,angle=270}\,\,\psfig{figure=GWWT_fig5h.ps,height=5.5cm,width=6.5cm,angle=270}}
\caption{Column-density plots of the evolution of the region around the 
primary star in Run A022, showing the development of spiral features. The 
spatial scale of the region is given in au, and the frame is centred on the 
primary star. The grey-scale bar gives the column-density in 
${\rm g}\,{\rm cm}^{-2}$, and the time is shown in Myr above each panel. 
Inflow occurs preferentially from the left and right and perturbs the disc 
around the primary star, exciting spiral instabilities.}
\label{fig:spiral}
\end{figure*}

\begin{figure*}
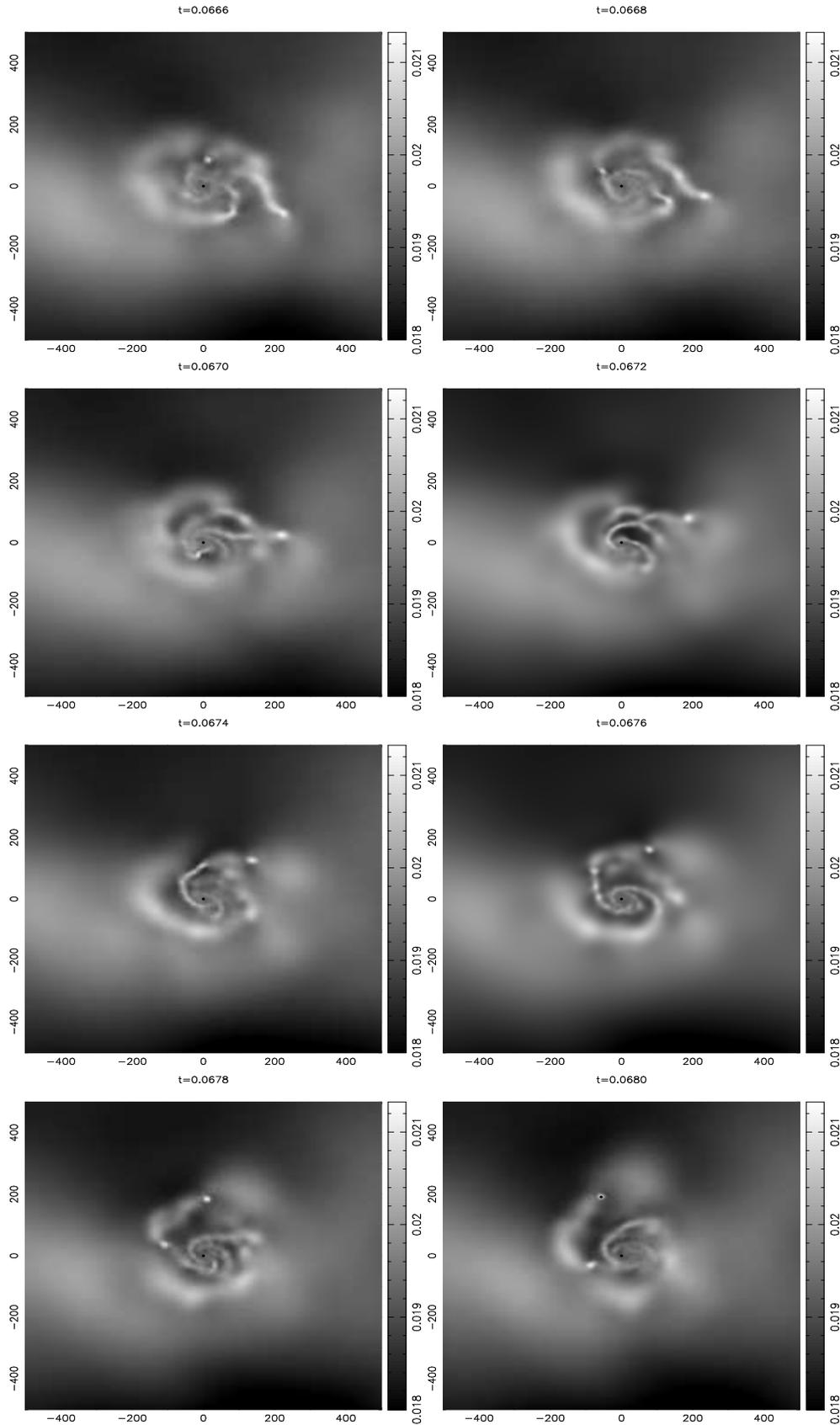

\centerline{\psfig{figure=GWWT_fig6a.ps,height=5.5cm,width=6.5cm,angle=270}\,\,\psfig{figure=GWWT_fig6b.ps,height=5.5cm,width=6.5cm,angle=270}}
\vspace{0.1cm}
\centerline{\psfig{figure=GWWT_fig6c.ps,height=5.5cm,width=6.5cm,angle=270}\,\,\psfig{figure=GWWT_fig6d.ps,height=5.5cm,width=6.5cm,angle=270}}
\vspace{0.1cm}
\centerline{\psfig{figure=GWWT_fig6e.ps,height=5.5cm,width=6.5cm,angle=270}\,\,\psfig{figure=GWWT_fig6f.ps,height=5.5cm,width=6.5cm,angle=270}}
\vspace{0.1cm}
\centerline{\psfig{figure=GWWT_fig6g.ps,height=5.5cm,width=6.5cm,angle=270}\,\,\psfig{figure=GWWT_fig6h.ps,height=5.5cm,width=6.5cm,angle=270}}
\caption{Column-density plots of the evolution of the region around
the primary star in Run A022 as it produces a second object. Labeling 
is as in Fig.~\ref{fig:spiral}. The spiral features that are shown 
forming in Fig.~\ref{fig:spiral} become self-gravitating and form 
knots along the arms. A sink condenses out of one of these knots in 
the last panel. This knot can be followed anti-clockwise starting from 
$(x,y) = (+225\,{\rm au},-100\,{\rm au})$ in the first panel.}
\label{fig:further}
\end{figure*}

The evolution of Run A022 is shown in Figs.~\ref{fig:spiral} and 
~\ref{fig:further}, as an illustration of these processes at work. 
In all frames the primary star is at the centre of co-ordinates. 
Fig.~\ref{fig:spiral} shows the lumpy accretion flow onto the 
primary star, concentrated preferentially through two streams 
on either side of the primary star. The lumpy inflow causes the 
disc to become unstable, forming spiral arms. Eventually one of 
these arms sweeps up sufficient material to detach and condense 
into a secondary object, as shown in Fig.~\ref{fig:further}; the 
knot that forms the secondary object is at $(x,y) 
\sim (+225{\rm au},-100{\rm au})$ in the first panel of Fig.~\ref{fig:further}, 
and orbits the primary anti-clockwise, 
ending up as a newly-formed sink at $(x,y) \sim (-80{\rm au},+200{\rm au})$ 
in the final panel of Fig.~\ref{fig:further}. We emphasise that this 
knot has already condensed out of the spiral arm that spawned it, 
by the time that it becomes a sink. A third sink is formed shortly 
after the illustrated sequence ends, from a knot that can be seen in 
the last three panels moving anti-clockwise from
$(x,y) \sim (-50{\rm au},+100{\rm au})$ to 
$(x,y) \sim (-50{\rm au},-50{\rm au})$.

Not all the knots that form in this way evolve into sinks. 
Some are destroyed by tidal interaction and/or merger with an 
existing sink. An example of this can be seen in Fig.~\ref{fig:further} 
where a dense knot at $(x,y) \sim (0{\rm au},+80{\rm au})$ in the 
first panel spirals into, merges with, the primary star.

The most significant factor in forming multiple objects appears to 
be the ability of the turbulent flows to create an extended overdense 
region in the vicinity of the primary star. The more material that 
is delivered into this region, the more objects that form. When an 
extended overdense region does not form around the primary star, 
the result is that no further objects form, e.g. runs A020, A021, 
A028 and A030.

Whilst this may appear similar to fragmentation of a disc 
formed from the collapse of a purely rotating cloud, there are 
significant differences in the
scenario outlined above.  Turbulence generates the angular momentum
required to
create a disc, but this angular momentum is provided by the turbulence
in the vicinity of the first object, and not from any bulk properties of the
core.  It is then the clumpy, inhomogeneous inflow from the
turbulent surroundings onto this disc that causes the disc to 
become unstable.  There is no correlation between the initial
angular momentum of the core and the number of objects that later
form.  The 3 cores with the lowest initial angular momentum (runs 
A013, A012 and A028)form 10, 4 and 1 objects , and the 3 cores
with the highest initial angular momentum (runs A019, A014 and A016) 
form cores with 8, 3 and 3 objects.  The amount of angular
momentum varies by a factor of 4.3 between these two extremes.  The
differences between runs A013 and A028 are solely due to the
different infall histories caused by the turbulent velocity field.

\subsection{Few-body interactions and ejections}

If more than two objects form in a core, the resulting 
${\cal N}$-body system is generally unstable (e.g. Valtonnen 
\& Mikkola 1991). The dynamics of unstable 
multiple systems are chaotic, and usually result in the ejection 
of low-mass members and the hardening of 
binaries (Anosova 1986; Sterzik \& Durisen 1998).

In our simulations, this dynamical phase usually ends about 0.10 to 
0.12 Myr after the start of the simulation, leaving an expanding halo 
of ejected objects and a central system containing between 2 and 4 
stars. Simulations that produce more than two objects show a high level 
of dynamical instability, and no systems remain with more than 4 
objects bound in the central region after $0.3\,{\rm Myr}\,$.

The timescale for dynamical evolution and ejection matches that 
given by Anosova (1986) who argues that systems decay on a timescale 
of order one hundred crossing times, i.e.
\begin{equation}
t_{\rm decay} \sim 0.17\,{\rm yr} 
\;\left( \frac{R}{\rm au} \right)^{3/2}
\;\left( \frac{M}{M_\odot} \right)^{-1/2}
\end{equation}
where $R$ is the scale-length of the system (here $\sim 200\,{\rm au}$) 
and $M$ is the total mass (here $\sim 2 M_\odot$), so we obtain 
$t_{\rm decay} \sim 0.03\,{\rm Myrs}\,$, which is a good fit to 
the decay timescales we observe in the simulations.

\begin{figure*}
\centerline{\psfig{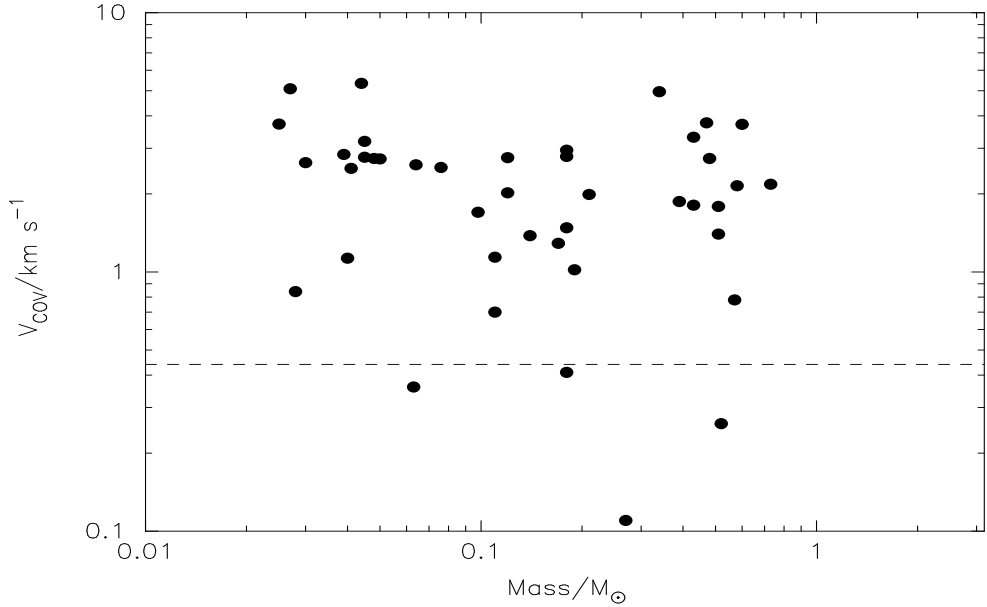}}
\caption{The velocities of ejected stars plotted against their masses. 
The dashed-line at $0.44\,{\rm km}\,{\rm s}^{-1}$ shows the initial escape 
speed from the core. Brown dwarves have a higher velocity than low-mass 
stars after being ejected.  The maximum mass of ejected stars is  
$\sim 0.5 M_{\odot}$.}
\label{fig:massvel}
\end{figure*}

Figure~\ref{fig:massvel} shows the relationship between the 
velocities of the objects formed (relative to the centre of 
mass of the core) and their masses. The initial escape velocity 
from a core is $\sim 0.44\,{\rm km}\,{\rm s}^{-1}$ and this is 
marked by the horizontal dashed-line on Fig.~\ref{fig:massvel}). 
There is a slight anti-correlation between ejection velocity 
and mass. Brown dwarves have a higher mean ejection velocity 
($\sim 2.9\,{\rm km}\,{\rm s}^{-1}$) than stars 
($\sim 2.0\,{\rm km}\,{\rm s}^{-1}$), but this correlation is 
not statistically very significant.

Dynamical interactions and ejections have been proposed by Reipurth 
\& Clarke (2001) (see also Bate et al. 2002a, Delgardo-Donate et al. 
2003) as a mechanism for the production of brown dwarves: 
stellar embryos are ejected from cores before they can accrete 
enough material to become hydrogen-burning stars. This appears to 
be the formation mechanism for all but one of the brown dwarfs formed in 
these simulations. Dynamical interactions eject 36 low-mass objects 
from our cores, and of 
these 14 have not accreted enough material to pass the 
hydrogen-burning limit at $0.08 M_{\odot}$. Only one brown dwarf 
is still bound in a core at $0.3\,{\rm Myr}\,$; that one is in
a binary with a $0.29 M_{\odot}$ star.

\subsection{The mass function}

\begin{figure*}
\centerline{\psfig{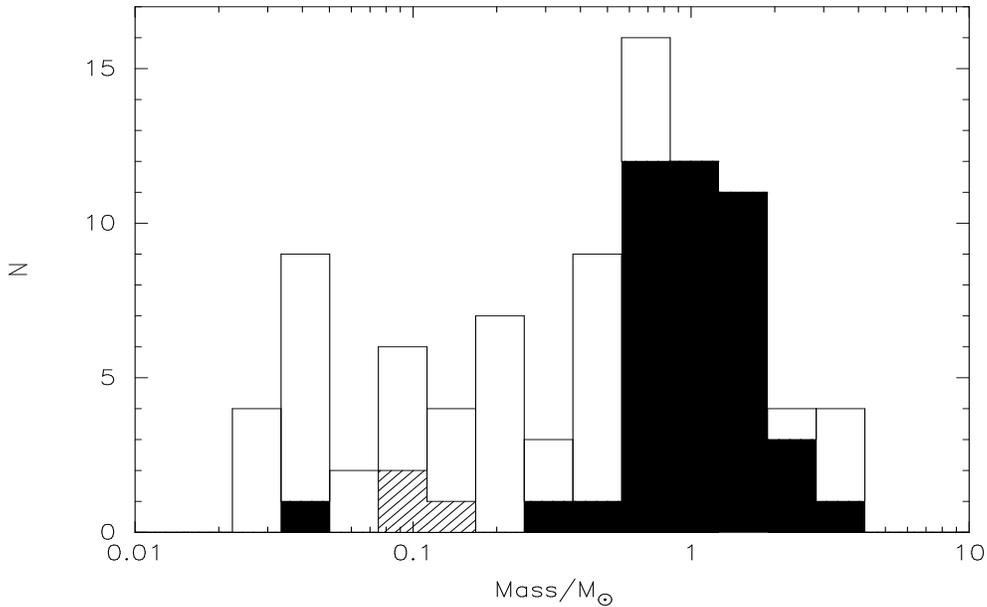}}
\caption{The mass function of stars created in the 20 low-turbulence 
simulations. Filled bins show objects in binary or multiple systems,
while open bins show single objects.  The three low-mass objects 
which are partially shaded are the three objects that formed late in 
Run A013; these objects are bound but highly unstable when the simulation 
is terminated at $0.3\,{\rm Myr}$, and their future evolution is 
therefore uncertain.}
\label{fig:imf}
\end{figure*}

Figure~\ref{fig:imf} shows the mass function of objects from all 
of the low-turbulence simulations at $0.3\,{\rm Myr}\,$. The 
filled portion of the histogram shows objects that are in multiple 
systems, while the open part shows single objects, and the hashed 
region shows the three low-mass stars which form late in Run A013; 
the final status of these three stars is unclear as the system is highly 
unstable when the simulation ends. The probability of ejection 
scales as $\sim M^{-1/3}$ (Anosova 1986) and ejected objects very 
seldom belong to multiple systems. Consequently all but one of the 
brown dwarfs and most of the low-mass stars are single and have been  
ejected (see also Fig.~\ref{fig:massvel}). The proportion of single 
stars decreases with increasing mass, because the longer a star 
remains in a core, the larger its mass grows, and the less likely 
it is to be ejected. 
 
The low-mass tail in the mass function below $\sim 0.5 M_{\odot}$ 
arises because secondary objects have difficulty growing beyond 
that mass before they are ejected. For example, the mean 
ejection timescale of $\sim 0.03\,{\rm Myr}$ multiplied by the 
mean initial accretion rate of $\sim 10^{-5} M_{\odot}\,{\rm yr}^{-1}$ 
gives a typical mass at ejection of $\sim 0.3 M_{\odot}$. Stars 
with final masses greater than this are likely to be part of 
a central multiple system, as only in the dense central region, 
where accretion is on-going, can the mass grow beyond 
$\sim 0.5 M_{\odot}$. This is shown in Fig.~\ref{fig:imf} by 
the high proportion of stars having $M > 0.5 M_{\odot}$ 
which are still in multiple systems at $0.3\,{\rm Myr}\,$.

The mass function shown in Fig.~\ref{fig:imf} should not be taken 
as a full initial mass function (IMF).  It represents the statistical 
output from only one mass of core, with only one level of turbulence. 
Figure~\ref{fig:imf} is apparently deficient in low-mass objects 
and brown dwarfs compared with the observed IMF (e.g. Kroupa 2002). 
Lower mass cores may well produce the smaller objects needed to 
populate this region, as their gas reservoir is smaller (e.g. 
Delgardo-Donate et al. 2003; Sterzik \& Durisen 2003). Alternatively, 
more turbulent cores may eject objects more rapidly, and hence with 
lower masses (e.g. Bate et al. 2002a,b; 2003). The true 
IMF is presumably a convolution of the object production from 
a distribution of cores masses, turbulence levels, and so on.  
For example, Delgardo-Donate et al. (2003) convolve the distribution 
of objects produced by one mass of core and one of level of turbulence 
with a core mass function, to produce an IMF.  What is clear from
these simulations is that the relationship between a core mass
spectrum and a stellar IMF is non-trivial and that core mass
spectra that do not resemble the stellar IMF (e.g. some of the core
mass spectra from Klessen 2001) may still produce a reasonable 
stellar IMF.

\subsection{Single stars}

Four simulations produce only single stars.  The turbulent velocity
fields in these simulations are such that an extended overdense 
region does not form. The core collapses in a  monolithic fashion 
similar to the zero-turbulence case (see Section 4.1). The only 
difference is that the accretion rate is somewhat lower, and the 
final mass at $0.3\,{\rm Myr}$ is therefore somewhat less than in the 
zero-turbulence case. For example, the primary star in Run A030 
only reaches a mass of $2.62 M_{\odot}$ by $0.3\,{\rm Myr}$, as 
compared with $\sim 3.75 M_{\odot}$ in the runs with no turbulence. 
This is because the extra support provided by turbulence slows down 
the rate of infall onto the central sink.

\subsection{Binary \& multiple properties}

Of the 91 objects formed, 45 remain bound in binary and multiple systems
at the end of the simulations ($0.3\,{\rm Myr}$). The properties 
of these systems are summarised in Table~\ref{tab:binary} and 
also in Appendix 1. 

A useful measure of the multiplicity of a stellar population is 
given by the companion star frequency
\begin{equation}
{\rm CSF} = \frac{B + 2T + 3Q + ...}{S + B + T + Q + ...}
\end{equation}
(Patience et al. 2002), where $S$ is the number of single stars, 
$B$ the number of binaries, $T$ the number of triples, etc. 
Summing over all the low-turbulence 
simulations and making no distinction between stars and brown 
dwarves, we have $S=46$, $B=7$, $T=6$ and $Q=2$ (plus one highly 
unstable quintuple system from Run A013), giving a net CSF of 0.47. 
Observations of older clusters and of the field give a CSF of 
$\sim 0.1$ (e.g. Duquennoy \& Mayor 1991; Patience et al. 2002), 
but in young star forming regions the CSF can be far higher at 0.3 
to 0.4 (Reipurth \& Zinnecker 1993; Reipurth 2000; Patience et al. 
2002).  

The observed CSF is also very dependent on the primary mass (see 
Sterzik \& Durisen 2003 and references therein). In our simulations, 
of the 41 objects having $M \leq 0.5 M_{\odot}$, only 6 are in 
multiples, giving a low-mass CSF of $\sim 0.1$; two of these are 
in a binary system whose components comprise a low-mass star and 
a brown dwarf, and three are in the highly unstable triple produced 
in Run A013. Of the 50 stars having $M \geq 0.5 M_{\odot}$, only 
11 are not in multiple systems (and 4 of these form as single stars), 
giving a high-mass CSF of $\sim 0.9$. This dramatic difference
between the multiplicities for low and high mass objects arises 
because the low-mass objects tend to have been ejected from the 
core early in their existence, whereas the high-mass stars stick 
around near the centre of mass of the core.

\begin{table}
\caption{The instantaneous properties of the binary and multiple 
systems formed in all of the low-turbulence simulations after 
$0.3\,{\rm Myr}\,$. The six columns give the run identifier, 
the primary masses, 
$M_P$, the secondary masses, $M_S$, the semi-major axes, $a$, 
the eccentricities, $e$, and the mass ratios, $q$, of the system. 
In higher-order multiples when one component is a binary, the 
primary mass is replaced by a 'B' and the mass ratio is omitted; 
for example, the triple system formed in run A014 consists of an 
inner binary with component masses 1.63 and $0.83 M_{\odot}$ and 
separation $12.8\,{\rm au}\,$, plus a third star of mass $1.56 
M_{\odot}$ orbiting the close binary at $90.6\,{\rm au}$).}
\label{tab:binary}
\begin{tabular}{lcclll}
\hline
&&&&&\\
Run & $M_P/M_{\odot}$ & $M_S/M_{\odot}$ & $a/$au & $e$ & $q$ \\
&&&&\\
\hline
&&&&&\\
A011 & 1.31 & 0.61 & 6.8  & 0.78 & 0.47 \\
A012 & 2.32 & 0.74 & 122  & 0.83 & 0.32 \\
A013 & 1.07 & 0.66 & 22.9 & 0.87 & 0.62 \\
A014 & 1.63 & 0.83 & 12.8 & 0.10 & 0.51 \\
     &  B   & 1.56 & 90.6 & 0.09 &      \\
A015 & 2.63 & 1.06 & 281  & 0.17 & 0.40 \\
A016 & 2.18 & 1.40 & 724  & 0.31 & 0.64 \\
A017 & 1.60 & 1.16 & 5.9  & 0.28 & 0.73 \\
     & B    & 0.64 & 62.2 & 0.07 &      \\
A018 & 1.09 & 1.03 & 4.8  & 0.84 & 0.94 \\
     & B    & 0.69 & 205  & 0.81 &      \\
A019 & 1.27 & 1.16 & 10.8 & 0.43 & 0.92 \\
     &  B   & 0.69 & 78.9 & 0.26 &      \\
A022 & 0.91 & 0.69 & 14.5 & 0.13 & 0.76 \\
     & 1.52 & 0.89 & 12.1 & 0.05 & 0.59 \\
     &  B   &  B   & 79.3 & 0.09 &      \\
A023 & 1.43 & 0.70 & 4.6  & 0.98 & 0.49 \\
     &  B   & 0.83 & 460  & 0.82 &      \\
A024 & 1.46 & 1.28 & 18.8 & 0.21 & 0.87 \\
A025 & 1.43 & 0.76 & 125  & 0.54 & 0.48 \\
A026 & 1.23 & 1.03 & 4.0  & 0.82 & 0.84 \\
     &  B   & 0.71 & 269  & 0.45 &      \\
A027 & 3.19 & 0.48 & 170  & 0.15 & 0.15 \\
A029 & 1.20 & 0.89 & 10.6 & 0.63 & 0.74 \\
     & 0.29 & 0.041& 37.6 & 0.33 & 0.14 \\
     &  B   &   B  & 989  & 0.60 &      \\
&&&&&\\
\hline
\end{tabular}
\end{table}

The properties of the multiple systems formed change rapidly 
during the early evolution of a core. The chaotic evolution 
of a few-body system, combined with high accretion rates and 
motion through a dense gas, mean that any binary properties 
are, at best, short-lived. Once a stable system of two to four 
bodies is established, the binary characteristics also tend to 
stabalise. All the values quoted in Table~\ref{tab:binary} are 
the instantaneous values at $0.3\,{\rm Myr}\,$, when the 
simulations were terminated. Generally the binary properties 
have been stable since 0.10 to $0.15\,{\rm Myr}\,$, but in 
simulations which form high-order multiples the orbital parameters 
$a$ and $e$ can change abruptly

For example, in Run A029 a binary system comprising a low-mass 
star and a brown dwarf encounters a high-mass central binary 
around $0.22\,{\rm Myr}\,$. The effect of this interaction is 
to harden the high-mass binary (its semi-major axis decreases 
from $a=61\,{\rm au}$ to $a=14\,{\rm au}$) and to throw the 
low-mass binary into a larger orbit ($a \sim 1000\,{\rm au}$) 
while also softening this binary (from $a \sim 30\,{\rm au}$ 
to $a \sim 40\,{\rm au}$). In Run A026, an unstable 4-body 
system is formed and a $0.65 M_{\odot}$ star is ejected, 
leaving the two most massive stars in a very close binary 
($a \sim 6\,{\rm au}$) with the third star orbiting at 
$\sim 270\,{\rm au}\,$.

The effect of ongoing accretion can be either to harden or 
to soften a binary. If the accretion is asymmetric, it can act 
to increase or decrease the velocity of the companions relative 
to each other, and hence change their orbital parameters. 
Even if the accretion is symmetric, the net effect will depend 
on whether the accreted gas has higher or lower specific angular momentum 
than the binary system (Turner et al. 1995; Whitworth et al. 1995; 
Bate \& Bonnell 1997).

For example, in Run A012 at $0.2\,{\rm Myr}\,$, the binary 
components have masses of $2.07$ and $0.60 M_{\odot}$ and 
their orbit has a semi-major axis of $18.4\,{\rm au}\,$. 
However, by $0.3\,{\rm Myr}\,$, the orbit has softened to $122\,{\rm au}$ 
and the masses have increased to $2.32$ and $0.74 M_{\odot}$; 
accretion of gas with high specific angular momentum has caused 
the binary to soften. Conversely, in Run A019, accretion of gas 
with low specific angular momentum increases the mass ratio of 
the binary from $q=0.75$ to $q=0.92$ and also hardens it from 
$a=16\,{\rm au}$ to $a=11\,{\rm au}\,$. 

\begin{figure*}
\centerline{\psfig{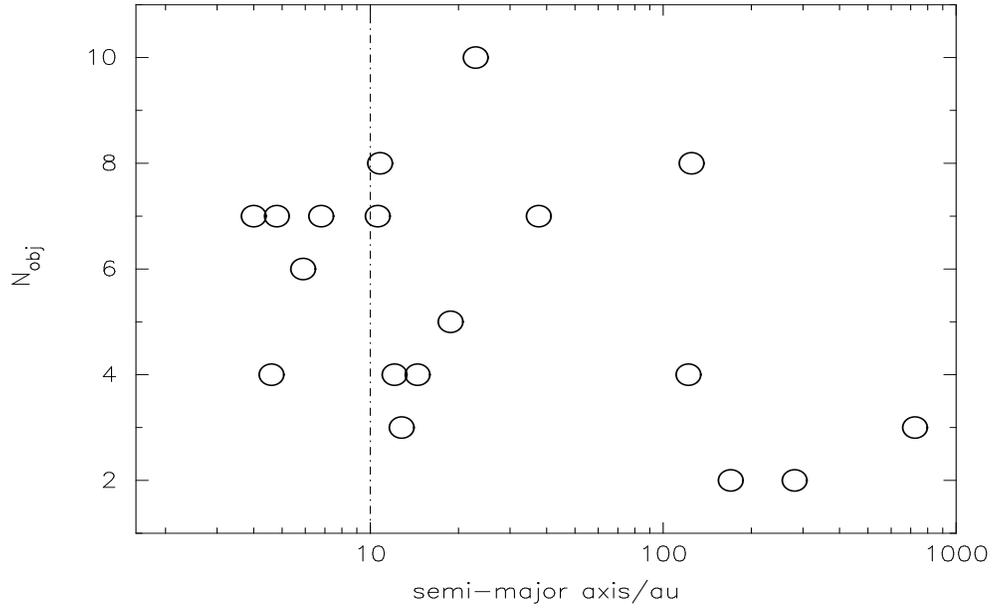}}
\caption{The total number of objects formed in a simulation $N_{\rm obj}$ 
plotted against the semi major axis of any binary systems formed in that 
simulation. Simulations which produce a large number of objects tend to 
produce harder binary systems.}
\label{fig:anobj}
\end{figure*}

While accretion can significantly affect the orbital separations of 
binaries, dynamical interactions appear to be the most important 
process generating (relatively) hard binaries in our simulations. 
This is shown in Fig.~\ref{fig:anobj}, where the semi-major axes of 
binaries are plotted against the total number of objects that form 
in a simulation. Of the 10 binaries in simulations that produce 
$\geq 5$ objects, 8 have $a < 30\,{\rm au}\,$, compared to only 4 
out of 8 binaries in simulations that produce $< 5$ objects. Both 
simulations that only produce 2 objects (and so never have any 
ejections) form wide binaries with $a > 100\,{\rm au}\,$.

It should be noted that for very close binary systems 
($a < 10\,{\rm au}$), we do not integrate the dynamics 
properly, due to the orbits being within the kernel-softened 
potential of the sinks. For this reason, the orbital parameters 
of close binaries are not robust. In particular, the 
eccentricities, and to a lesser extent the semi-major axes, 
are reduced by this numerical artifact

\begin{figure*}
\centerline{\psfig{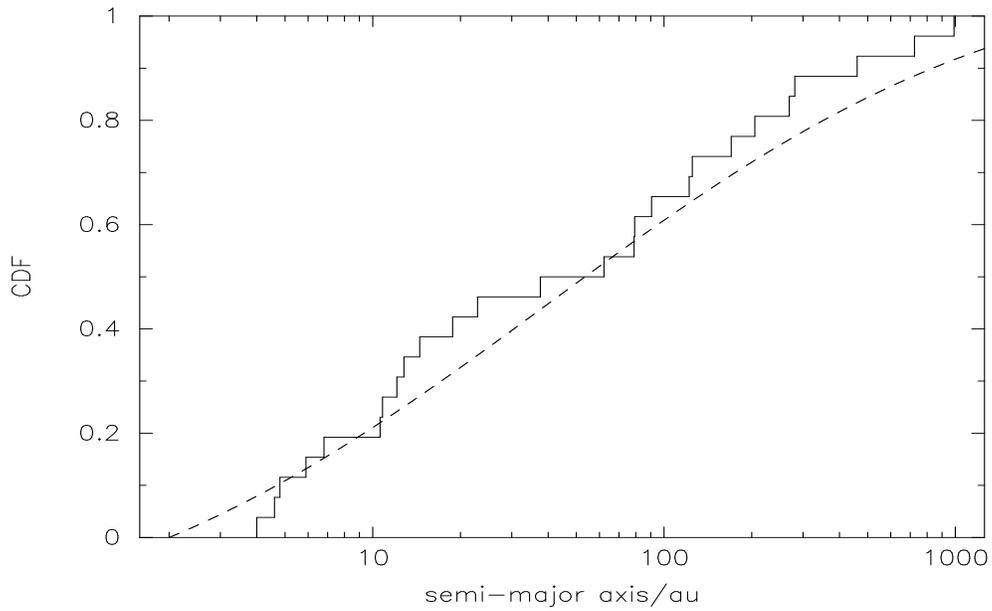}}
\caption{The cumulative distribution function of semi-major axes for the 
systems formed in the 20 low-turbulence simulations. The dashed line is the 
Gaussian fit to the G-dwarf period distribution from Duquennoy \&
Mayor (1991), adjusted to be a fit to the semi-major axis distribution by 
adopting a system mass of $M_\odot$. A KS test shows the two distributions 
to be compatible.}
\label{fig:semimajor}
\end{figure*}

Figure~\ref{fig:semimajor} shows the cumulative distribution of 
semi-major axes from the simulations at $0.3\,{\rm Myr}\,$. The 
majority of binary systems have semi-major axes 
$a \sim 10\;{\rm to}\;20\,{\rm au}\,$, and the high-$a$ tail 
consists mainly of triple and quadruple systems. Also plotted 
in Fig.~\ref{fig:semimajor} is a dashed-line giving the Duquennoy 
\& Mayor (1991) Gaussian fit to their period distribution, 
converted to give a distribution of semi-major axis by assuming 
a total system mass of $1 M_\odot$. The distribution of semi-major 
axes from our simulations is consistent with that from Duquennoy 
\& Mayor, in the sense that a KS test does not reject the sample 
of 26 semi-major axes as being drawn from the Duquennoy \& Mayor 
distribution. The distribution of semi-major axes is unlikely to 
change significantly due to the subsequent evolution of orbital 
parameters, because any hardening will be counter-balanced by softening.

\begin{figure*}
\centerline{\psfig{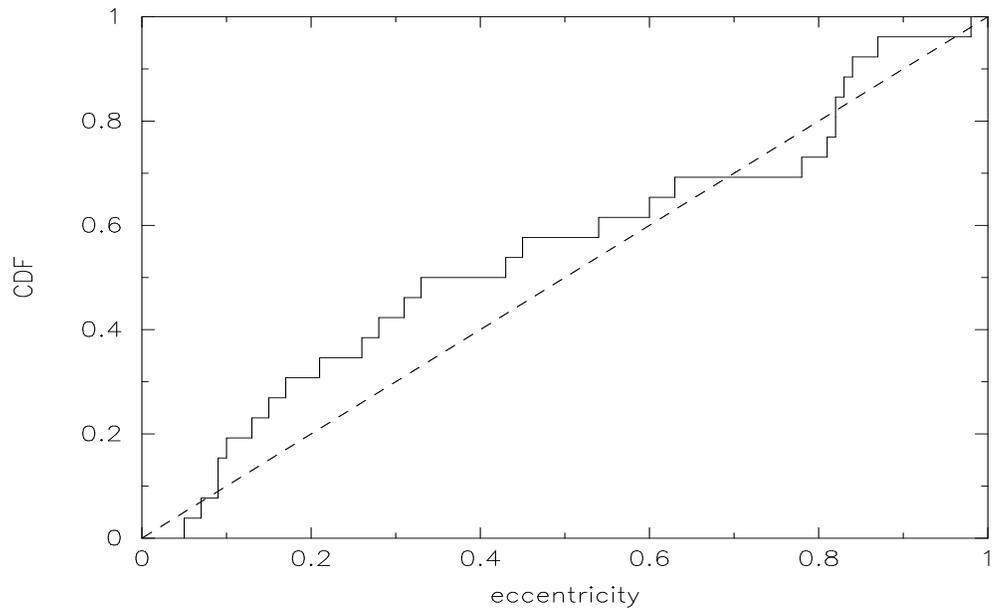}}
\caption{The cumulative distribution function of eccentricities for the 
systems formed in the 20 low-turbulence simulations. The dashed line is the 
linear fit from Duquennoy \& Mayor (1991), and is compatible with the simulations.}
\label{fig:ellipticities}
\end{figure*}

Figure~\ref{fig:ellipticities} shows the cumulative distribution 
function of the eccentricity of multiple systems.  Duquennoy \& Mayor 
(1991) observed a roughly linear cumulative distribution 
for their long-period sample ($P > 1000\,{\rm days}$) sample, as 
marked by the dashed-line, and this is consistent with our results. 
Duquennoy \& Mayor (1991) and Mathieu (1994) found that the spread 
of eccentricities increases with increasing separation, but this is 
only significant if short-period systems ($P < 1000\,{\rm days}$; 
$a <$ a few au) are included. We find the same trend in our 
results, but it is not statistically significant.

\begin{figure*}
$T=0.2$ Myr \\
\centerline{\psfig{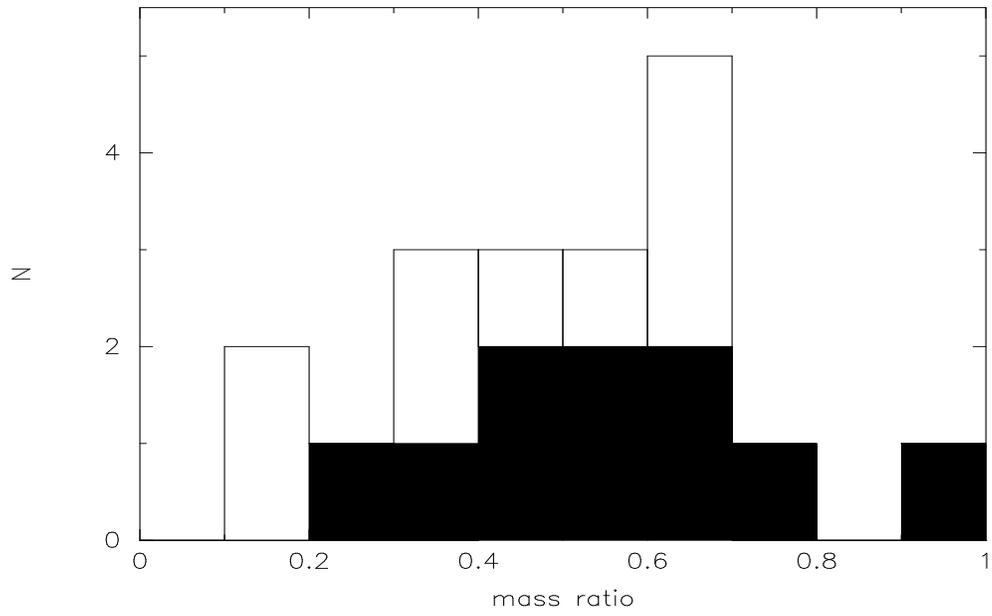}}
$T=0.3$ Myr \\
\centerline{\psfig{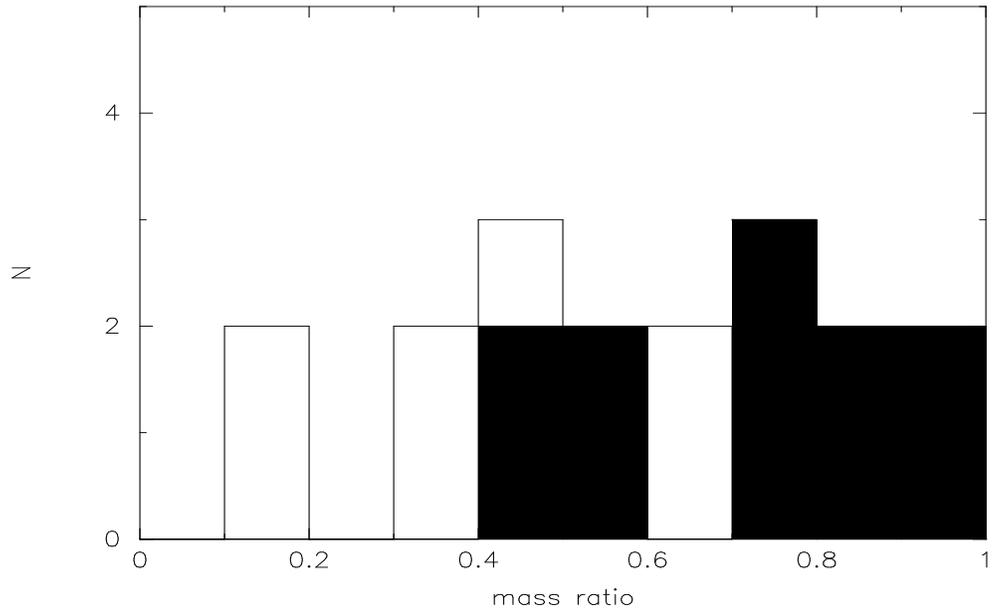}}
\caption{The distribution of binary mass ratios at $0.20\,{\rm Myr}$
and $0.30\,{\rm Myr}$ for the binary systems that form in the 
20 low-turbulence simulations. The filled bins show the 
mass ratios for close binaries ($a < 20\,{\rm au}$), and the empty bins show 
the mass ratios for wide binaries ($a > 20\,{\rm au}$). It should be
noted that one binary system at $0.20\,{\rm Myr}$ has been disrupted
by $0.30\,{\rm Myr}$, resulting in one less pair in the second histogram.}
\label{fig:massratios}
\end{figure*}

Figure~\ref{fig:massratios} shows the distribution of mass 
ratios of the binary systems in all of the simulations, at 
$0.2\,{\rm Myr}\,$ and at $0.3\,{\rm Myr}\,$. In both cases, the solid 
histograms show hard binary systems with $a < 20\,{\rm au}$, while 
the open histograms are for wider systems.  

The mass ratios of systems are the one property that still 
evolves significantly in all simulations after $0.2\,{\rm Myr}$ as is
clearly shown in Fig.~\ref{fig:massratios}. 
During evolution, there is a tendency for mass ratios in binary 
systems to become closer to unity (cf. Turner et al. 1995; 
Whitworth et al. 1995; Bate \& Bonnell 1997). This is because 
the lighter companion is more able to accrete high angular 
momentum material, and its wide orbit through the disc gives it 
more chance to accrete material. For example, the close binary 
in Run A019 has a mass ratio of $q=0.75$ at $0.2\,{\rm Myr}$ 
($M_P = 1.06$ and $M_S = 0.80 M_{\odot}$), but by $0.3\,{\rm Myr}$ 
the mass ratio has risen to $q=0.92$ as the primary mass increases 
to $1.27 M_{\odot}$ and the secondary mass increases to 
$1.16 M_{\odot}$.  As noted above this accretion also hardens the 
binary. Even in cases where the accretion rate onto both stars is 
very similar, the mass ratio increases because the proportional change 
in the secondary mass is greater. This results in a correlation of 
mass ratio with semi-major axis by $0.3\,{\rm Myr}\,$, with all 12 
close systems ($a < 20\,{\rm au}$) having $q > 0.45$ and 5 out of 
6 wide binaries ($a > 20\,{\rm au}$) having $q < 0.5$ (the one 
binary in the latter subset with $q > 0.5$ only has $q = 0.63$).

The distribution of mass ratios in the simulations at $0.3\,{\rm Myr}$ 
is close to the observed distribution, i.e. roughly linear between $q=0$ 
and $q=1$ (Mazeh et al. 1992, after correcting for selection 
effects on close binaries in the sample of Duquennoy \& Mayor 
1991).  However, at $0.3\,{\rm Myr}$ the simulations produce 
fewer binaries with $q \sim 1$ than are observed by White \& 
Ghez (2001) in Taurus-Auriga. White \& Ghez (2001) do 
observe a correlation between binary separation and mass ratio in 
Taurus-Auriga, similar to that seen in Fig.~\ref{fig:massratios}.

\section{Discussion}

The main mechanism for the formation of multiple systems in 
dense low-turbulence cores appears to be the fragmentation of 
circumstellar discs. The first star to form attracts a large 
circumstellar disc, which is then perturbed and destabalised 
by non-uniform infall, and fragments to form multiple systems.  
These discs do not form due to the bulk angular momentum of 
the core, but rather because of  local angular momentum in the region 
in which the
first object forms.  There is no correlation between the total angular
momentum of a core and the number of objects that form in it.  It is the
detailed local structure of the turbulent velocity field that
determines how many objects form, by first forming a circumstellar disc
on scales of a few hundred au, and then by perturbing that disc due to 
inhomogeneous infall onto the disc. 

Multiple formation occurs on the scale of a few hundred au, and 
the advent of high resolution sub-mm observations with instruments 
like ALMA will make it possible to probe this regime in unprecedented 
detail. The detection of highly unstable dense discs with 
non-axisymmetric instabilities would be a powerful confirmation of 
the relevance of the processes we have described above to the 
formation of multiple systems.

In a low-$N$ dynamical system it is unlikely that the orbits are 
stable and so dynamical interactions preferentially eject the low-mass 
members of a system (Anasova 1986).  Many objects are ejected at 
relatively high velocity before they are able to accrete enough 
material to become stars, thereby producing brown dwarfs. We agree 
with the hypothesis of Reipurth \& Clarke (2001) and Bate et al. 
(2002a) that an effective mechanism for the formation of brown 
dwarves is the ejection of protostellar embryos from their natal 
cores. Such a mechanism also explains the low binarity of brown 
dwarfs. In our simulations, only one brown dwarf is present in a 
multiple system.  

The average ejection velocity of low mass objects is 
$\sim 2\,{\rm km}\,{\rm s}^{-1}$, but brown dwarves have a 
slightly larger average ejection velocity than low-mass 
stars. Also, no binary systems are ejected 
from cores. Thus, dynamical ejections lead to a population of 
low-mass objects with low binarity. The higher velocity dispersion 
of this population should result in mass segregation in young 
clusters on a relatively short timescale. Indeed, the average 
ejection velocity of brown dwarfs ($\sim 2$ km s$^{-1}$) exceeds the
escape velocity from many small clusters. Thus, in small clusters 
of only a few Myrs, we would expect the spatial distribution of 
brown dwarves to be significantly larger than that of stars; and 
in older clusters the brown dwarf population may be significantly 
depleted by dispersion.

The companion star frequency (CSF) over our ensemble is 
$\sim 0.5$. This is far higher than in the field (e.g. Duquennoy 
\& Mayor 1991), but in good agreement with the high CSF observed 
for young clusters and T Tauri stars of 0.3 to 0.4 (Patience et al. 
2002), especially given that we do not have the selection 
effects which may have reduced the observed CSFs. Presumably, over time 
the CSF of a young stellar population  is altered by the disruption 
of multiple systems, reducing high initial CSFs to the field value 
(e.g. Kroupa 1995a, b).

The CSF is very dependent upon mass, with low mass objects 
($< 0.5 M_{\odot}$) having a CSF of only $\sim 0.1$, while 
more massive stars have a CSF of $\sim 0.9\,$. The huge 
difference between these two CSFs is due to the divide 
between low-mass systems which, almost by definition, are 
ejected from the natal core at an early stage in their 
growth, and high-mass systems which stick around in the 
centre of the natal core (interacting with the gas reservoir 
and with other stars) for a long time. This dependence of the 
CSF on mass is in the same sense as the observed decline of 
CSF with decreasing mass (Patience et al. 
2002; Sterzik \& Durisen 2003 and references therein).

The binary and higher multiple systems in our simulations have 
properties consistent with observations of pre-Main Sequence 
and Main Sequence binaries. The distribution of orbital 
separations is wide, between 4 and $1,000\,{\rm au}\,$, with 
a median semi-major axis of $\sim 30\,{\rm au}$. The 
distribution is consistent with a Gaussian fit to Duquennoy \& 
Mayor's (1991) sample of Main Sequence G-dwarves. Close binaries 
are formed mainly by dynamical hardening of wider binaries. 
The distribution of orbital separations is then further modified 
by accretion, which can both harden or soften existing binaries 
(Whitworth et al. 1995; Bate \& Bonnell 1997). The distribution 
of eccentricities is also consistent with the observations of 
Duquennoy \& Mayor (1991).

Mass ratios are the one binary property that evolves significantly 
in all of the simulations after $0.2\,{\rm Myr}\,$. At 
$0.2\,{\rm Myr}\,$, the mass ratios show a wide distribution with 
a peak in the range from $q = 0.5$ to $q = 0.7$.  As close binaries 
evolve, their mass ratios tend towards higher values as (a) the 
lower-mass companion is more able to accrete material due to its 
higher angular momentum and location in the disc, and (b) even with 
similar accretion rates, the proportional increase in the mass of 
the secondary is larger (Whitworth et al. 1995; Bate \& Bonnell 
1997). By $0.3\,{\rm Myr}\,$, the distribution of mass ratios 
shows a roughly linear rise with $q$, consistent with the 
observations of local G-dwarfs (Duquennoy \& Mayor 1991; Mahez et 
al. 1992). Close binaries have higher mass ratios than wide binaries; 
viz. all binaries with $a < 20\,{\rm au}$ have $q > 0.4$ by
$0.3\,{\rm Myr}\,$, whereas all but one of the binaries with 
$a > 20\,{\rm au}$ have $q < 0.5\,$. 

The ensemble of simulations presented in this paper represents only 
a single point in an extended parameter space. In future 
papers in this series we will examine the effect of different 
levels of turbulence on cores, as well as the effects of the power 
spectrum of the turbulence and the shape and structure of 
cores on star formation.  We will also examine cores with different masses.

\section{Conclusions}

We have presented an ensemble of simulations of star formation in 
turbulent dense cores, using initial conditions based on observation. 
The cores have an initial density profile with a flat $5,000\,{\rm au}$ 
central region (the kernel) and an $\approx 1/r^4$ fall-off beyond this out to 
$50,000\,{\rm au}$ (the envelope). The central density is 
$3 \times 10^{-18}\,{\rm g}\,{\rm cm}^{-3}$, and the total core mass 
is $5.4 M_{\odot}$. The initial ratio of thermal to gravitational 
energy is $\alpha_{\rm therm} = 0.45$.

Without turbulence, a single, central star forms, as would be expected 
from analytical studies (e.g. Whitworth \& Ward-Thompson 2001). We 
then add low levels of turbulence with a low ratio of turbulent to 
gravitational energy of $\alpha_{\rm turb} = 0.05$. The results can 
be summarised as follows.

\begin{itemize}

\item{This low level of turbulence results in the formation 
of multiple objects in 80\% of cores.}

\item{The number of objects that forms in a turbulent core depends 
in a chaotic way on the details of the turbulent velocity
field. Between 1 and 10 objects form in each of our 20 
simulations, with an average of 4.55 objects per core.}

\item{Close binaries ($a < 20\,{\rm au}$) are formed mainly by 
the hardening of wider binaries due to dynamical interactions.}

\item{The multiple systems that form in these simulations have 
semi-major axes, eccentricities and mass ratios compatible 
with the observed distributions.}

\item{Low-mass objects ($< 0.5 M_{\odot}$) are often formed and then 
ejected from cores by dynamical interactions. This leads to a
low-mass population with low binarity and high velocity dispersion.}

\end{itemize}

In summary, turbulence is a fundamental ingredient of core physics, 
and leads to the production of abundant multiple systems. The 
resulting systems have properties similar to those that are observed. 
The levels of turbulence required to produce multiple systems are 
very low, similar to those observed in more quiescent isolated cores. 
Since the fragmentation of a turbulent core is a chaotic process, it 
is essential to perform a statistically significant ensemble of 
simulations in order to parametrize the properties of the stars which 
form from cores of different mass, different levels of turbulence, 
etc.


\begin{acknowledgements}

SPG gratefully acknowledges the support of PPARC grant 
PPA/G/S/1998/00623. We have extensively used the Beowulf 
cluster of the gravitational waves group at Cardiff 
and thank B.\,Sathyprakash and R.\,Balasubramanian for allowing 
us access. We would like to thank Matthew Bate for helpful 
discussions and for providing the code used to generate the 
initial turbulent velocity field. SPG would also like to
thank Dave Nutter and Jason Kirk for explaining the 
observations of dense cores.  Thanks also go to the referee, Ralf
Klessen, for helpful comments and suggestions.

\end{acknowledgements}

\section{Appendix}

In this appendix we describe in more detail the evolution of each 
of the 20 low-turbulence simulations. In all cases, unless otherwise 
stated, the masses quoted here are the {\em final} masses of the stars 
or brown dwarves at $0.3\,{\rm Myr}\,$.

\subsection{A011}

Seven objects form in this simulation. The first, a $1.31
M_{\odot}$ star forms at $0.053\,{\rm Myr}\,$, and the other 
six (including two brown dwarfs) form between 0.070 and 
$0.073\,{\rm Myr}\,$. Four objects are ejected, and the 
remaining three stabalise by $0.12\,{\rm Myr}$ 
into a close binary (component masses 1.31 and $0.52M_{\odot}$, 
semi-major axis $a=9.8\,{\rm au}$) and a third relatively high-mass 
star ($0.61 M_{\odot}$) in a distant orbit ($a \sim 800\,{\rm au}$) 
around this binary. As the system evolves, the central binary 
hardens and the more distant companion becomes only very loosely 
bound with an orbit of a few thousand au. Two other stars, with 
masses 0.27 and $0.12 M_{\odot}$ are still in the core at 
$0.3\,{\rm Myr}\,$, but both are single.

\subsection{A012}

This simulation produces four stars, with masses 2.32, 0.74, 0.48 
and $0.18 M_{\odot}$. The first star forms after $0.056\,{\rm Myr}\,$, 
and the other three form in a short burst after $\sim 0.08\,{\rm Myr}\,$.  At $0.15\,{\rm Myr}$ the two most massive objects form a close, 
eccentric binary with $a \sim 18\,{\rm au}$ and $e \sim 0.75$, but by 
$0.3\,{\rm Myr}$ accretion has significantly softened this binary 
and $a \sim 122\,{\rm au}\,$.

\subsection{A013}

Five stars and one brown dwarf form early in this simulation. 
The primary star forms at $0.052\,{\rm Myr}$ and eventually 
acquires a mass of $1.07 M_{\odot}$. The other four stars and the 
brown dwarf form in a burst between 0.067 and $0.071\,{\rm Myr}\,$. 
By $0.20\,{\rm Myr}$, the primary star has paired up with a 
$0.66 M_{\odot}$ star to form a close binary with $a=23\,{\rm au}$ 
and $e=0.87$.  

An unusual second burst of star formation occurs between 0.26 and 
$0.28\,{\rm Myr}$, forming another four objects; by $0.3\,{\rm Myr}$, 
three have accreted enough material to become stars, and the fourth 
has been ejected as a brown dwarf. The three new stars form a very 
close, but highly unstable, triple, which is bound into a 
quintuple with the main binary system. However this quintuple is 
not expected to survive, and at least one of its low-mass components 
is likely to be ejected.

\subsection{A014}

Three stars form, with masses 1.63, 1.56 and $0.83 M_{\odot}$, 
the first after $0.052\,{\rm Myr}$, and the others at 
$\sim 0.075\,{\rm Myr}\,$. All three stars remain bound in a 
triple system; the 1.63 and $0.83 M_{\odot}$ stars are in a 
central close binary with $a=13\,{\rm au}$ and $e=0.10$, while 
the $1.56 M_{\odot}$ star is in an orbit with $a \sim 90\,{\rm au}$ 
and $e=0.09$ around this binary. The outer star in the triple has 
been able to grow significantly faster than the central stars as, 
unlike those stars, it does not have to share material falling 
into its gravitational influence.

\subsection{A015}

Only two stars are formed, a $2.63 M_{\odot}$ star after $0.057\,{\rm Myr}$,
followed by a $1.06 M_{\odot}$ star at $0.065\,{\rm Myr}$. These stars 
form and remain in a wide binary with $a = 280\,{\rm au}$.

\subsection{A016}

Two stars are formed early, a $2.18  M_{\odot}$ star after 
$0.057\,{\rm Myr}$, and a $1.40 M_{\odot}$ star after 
$0.072\,{\rm Myr}$. The evolution is very similar to A015 
until, at $0.178\,{\rm Myr}$, a third object forms and is 
immediately ejected as a brown dwarf.

\subsection{A017}

Six stars are formed, the first at $0.051\,{\rm Myr}$ and the 
other five in a late burst around $0.096\,{\rm Myr}$. After 
three stars have been ejected, a massive triple system remains, 
comprising a close binary (component masses 1.60 and $1.16 M_\odot$, 
semi-major axis $a=6\,{\rm au}$) with a third star ($0.64 M_{\odot}$) 
in a wide orbit with $a \sim 60\,{\rm au}$ around this binary.

\subsection{A018}

Five stars and two brown dwarfs are formed, the first star at 
$0.055\,{\rm Myr}$, the other four stars and one brown dwarf 
between 0.065 to $0.071\,{\rm Myr}$; the second brown dwarf 
forms at $0.10\,{\rm Myr}$ and is immediately ejected. Three 
of the stars form a triple with component masses 1.09, 1.03 
and $0.69 M_{\odot}$, but it is very unstable. At 
$0.20\,{\rm Myr}$ the two most massive stars are in a close 
binary, with the $0.69 M_{\odot}$ star in a more distant orbit. 
Then at $0.21\,{\rm Myr}$ there is an interaction which swaps 
the $0.69 M_{\odot}$ star with one of the components of the close 
binary, and at around $0.26\,{\rm Myr}$ they are swapped 
back.  In addition there is a $0.58 M_{\odot}$ star very loosely 
bound to the triple system.

\subsection{A019}

This simulation produces eight objects, five stars and three 
brown dwarves. The first object forms after $0.057\,{\rm Myr}$, 
followed by two bursts, the first of which forms four objects at 
$\sim 0.064\,{\rm Myr}$, and the second of which forms three objects 
at $\sim 0.090\,{\rm Myr}$. By the end of the simulation there is 
a massive triple comprising a close binary (component masses 1.27 
and $1.16 M_\odot$, semi-major axis $a=11\,{\rm au}$, eccentricity 
$e=0.43$) with a third star ($0.69 M_{\odot}$) in a wide orbit 
($a \sim 80\,{\rm au}$) around this binary.

\subsection{A020}

A single star is formed at $0.055\,{\rm Myr}$ and reaches a mass 
of $3.63 M_{\odot}$ by $0.3\,{\rm Myr}$.

\subsection{A021}

A single star is formed at $0.054\,{\rm Myr}$ and reaches a mass 
of $3.69 M_{\odot}$ by $0.3\,{\rm Myr}$.

\subsection{A022}

Four stars are formed in this simulation, the first after 
$0.057\,{\rm Myr}$, and the other three in a burst around 
$0.072\,{\rm Myr}$. All four stars remain bound at 
$0.3\,{\rm Myr}$ as an hierarchical quadruple consisting of a 
pair of close binary systems. The more massive binary comprises 
1.52 and $0.89 M_{\odot}$ stars in an orbit with $a=15\,{\rm au}$ 
and $e=0.13$. The less massive binary comprises 0.91 and 
$0.69 M_{\odot}$ stars in an orbit with $a=12\,{\rm au}$ and 
$e=0.05$. The two binary systems are in a wide orbit around 
each other with $a \sim 80\,{\rm au}$ and $e \sim 0.09$.

\subsection{A023}

Four stars form in this simulation. The primary star forms at 
$0.055\,{\rm Myr}$ and eventually acquires a mass of 
$1.43 M_{\odot}$.  The remaining three stars form between 0.069 
and $0.079\,{\rm Myr}$. By $0.3\,{\rm Myr}$ there is a massive 
triple comprising a close binary (component masses 1.43 and 
$0.70 M_\odot$, semi-major axis $a=5\,{\rm au}$) with a third 
star ($0.83 M_{\odot}$) in a wide orbit around this binary. The 
fourth star has $0.60 M_{\odot}$ and has been ejected with a 
large disc early in the interaction process.  

\subsection{A024}

Five stars are formed, the first at $0.053\,{\rm Myr}$, and the 
rest between 0.070 and $0.076\,{\rm Myr}$. Three of the stars 
are ejected, leaving a close binary with component masses $1.46$ 
and $1.28 M_{\odot}$, semi-major axis $a=19\,{\rm au}$ and 
eccentricity $e=0.21$. Unusually, the most massive star at the
end of the simulation is the second object to form, rather than 
the first as in all other simulations.

\subsection{A025}

Eight objects are formed in this simulation, including three brown
dwarves. The primary star forms at $0.053\,{\rm Myr}$, and five 
further objects, including the three brown dwarves, form in a burst 
around $0.064\,{\rm Myr}$; the final two stars form at 
$0.09\,{\rm Myr}$. Five objects are ejected almost immediately, 
leaving a triple system, which, despite the ejections is very loose; 
the main binary has $a \approx 220\,{\rm au}$. After 
$0.17\,{\rm Myr}$ the third component of the triple is ejected and 
the remaining binary is somewhat hardened to $a \sim 125\,{\rm au}$.  

\subsection{A026}

Six stars and one brown dwarf are formed. The primary star forms 
at $0.055\,{\rm Myr}$, and four further stars and the brown dwarf 
form in a burst around $0.071\,{\rm Myr}$; the final star forms at 
$0.092\,{\rm Myr}$. Two stars and the brown dwarf are ejected, 
leaving two close binaries in an hierarchical quadruple. The more 
massive binary comprises 1.03 and $0.71 M_{\odot}$ stars in an 
orbit with $a=15\,{\rm au}$. The less massive binary comprises 
0.73 and $0.71 M_{\odot}$ stars in an orbit with $a=16\,{\rm au}$. 
The two binary systems are in a wider orbit around each other with 
$a \sim 180\,{\rm au}$ and $e \sim 0.09$. However, this system is 
unstable, and at $0.24\,{\rm Myr}$ the $0.73 M_{\odot}$ star is 
ejected in a 4-body encounter, leaving the more massive binary 
hardened to $a = 4\,{\rm au}$ (below our ability to resolve the 
dynamics properly) and the $0.71 M_{\odot}$ star in a $270\,{\rm au}$ 
orbit around this binary, i.e. an hierarchical triple.

\subsection{A027}

Only two stars are formed in this simulation. The primary star 
is formed at $0.057\,{\rm Myr}$ and ends up with $3.19 M_{\odot}$. 
The secondary star forms at $0.142\,{\rm Myr}$ and ends up with 
$0.48 M_{\odot}$. They form a wide binary system with 
$a=170\,{\rm au}$ and $e=0.15$. The large mass ratio is due to 
the very late formation time of the secondary star and its 
distance from the primary.

\subsection{A028}

A single star is formed at $0.055\,{\rm Myr}$ and reaches a mass 
of $3.35 M_{\odot}$ by $0.3\,{\rm Myr}$.

\subsection{A029}

A total of seven objects is formed, the first at 
$0.057\,{\rm Myr}$, another five in a burst around 
$0.064\,{\rm Myr}$, and a final object at $0.081\,{\rm Myr}$. 
Three stars are ejected, leaving two close binary systems in 
an hierarchical quadruple. The more massive binary comprises 
1.20 and $0.89 M_{\odot}$ stars in an orbit with 
$a=11\,{\rm au}$ and $e = 0.63$. The less massive binary comprises 
0.29 and $0.041 M_{\odot}$ stars in an orbit with 
$a=38\,{\rm au}$ and $e = 0.33$; this is the only example of a 
brown dwarf in a binary or multiple system in all our simulations.   
The two binary systems are in a very wide orbit around each other 
with $a \sim 1500\,{\rm au}$.

\subsection{A030}

A single star is formed at $0.055\,{\rm Myr}$ and reaches a mass 
of $2.62 M_{\odot}$ by $0.3\,{\rm Myr}$.

\newpage

\newpage

\newpage

\newpage

\newpage

\newpage

\newpage

\newpage

\newpage

\newpage

\newpage

\newpage

\end{document}